  \documentclass{aa}  
  \usepackage{chngcntr}
  \usepackage{upgreek}
  \usepackage[overload]{empheq}
  \usepackage{subcaption} 
  \usepackage{tabularx} 
  \usepackage{booktabs}                     
  \usepackage{float}
  \usepackage{epstopdf}
  \usepackage{longtable}
  \usepackage[export]{adjustbox}
  \usepackage{caption}
  \usepackage{color,soul}
  \usepackage{amsmath}
  \usepackage{nameref}
  \usepackage{varioref}
  \usepackage{xcolor}
  \hypersetup{
      colorlinks,
      linkcolor={red!50!black},
      citecolor={blue!80!black},
      urlcolor={blue!80!black}
  }
  \usepackage{textcomp}
  \usepackage{cleveref}
  \usepackage[utf8]{inputenc}
  \usepackage{natbib}
  \bibpunct{(}{)}{;}{a}{}{,} 
  \counterwithout{figure}{section}
  \usepackage{cases}
  \usepackage{graphicx}
  \usepackage{placeins}
  \usepackage{upgreek}
  \usepackage{amsmath}	
  \usepackage{txfonts}
  \newcommand{\CS}{CS~$J$~=~7--6~}
  \newcommand{\HthirtCN}{H$^{13}$CN~$J$~=~4--3~}
  \newcommand{\HCN}{HCN~$J$~=~4--3~}
  \newcommand{\HCO}{HCO$^{+}$~$J$~=~4--3~}

\begin{document}

     \title{The organic chemistry in the innermost, infalling envelope of the Class~0 protostar L483}

     \author{Steffen K. Jacobsen
       \inst{\ref{starplan}}
            \and
            Jes K. J\o rgensen\inst{\ref{starplan}} 
            \and James Di Francesco\inst{\ref{James}} \and Neal~J.~Evans II\inst{\ref{Texas}, \ref{Neal_2}, \ref{Choi}} \and Minho~Choi\inst{\ref{Choi}} \and Jeong-Eun~Lee\inst{\ref{Jeong}}
            }

     \institute{Niels Bohr Institute \& Centre for Star and Planet Formation, University of Copenhagen, {\O}ster Voldgade 5--7, DK-1350 Copenhagen K., Denmark.
     \label{starplan}
          \and
     NRC Herzberg Astronomy and Astrophysics, 5071 West Saanich Road, Victoria, BC, V9E 2E7, Canada.
     \label{James}
     \and
     Department of Astronomy, The University of Texas at Austin, Austin, TX 78712, USA.
     \label{Texas}
     \and
     Humanitas College, Global Campus, Kyung Hee University, Yongin-shi 17104, Korea.
     \label{Neal_2}
     \and
     Korea Astronomy and Space Science Institute, 776 Daedeokdae-ro, Yuseong-gu, Daejeon 34055, Korea. 
     \label{Choi}
     \and
     School of Space Research, Kyung Hee University, 1732, Deogyeong-Daero, Giheung-gu, Yongin-shi, Gyunggi-do 17104, Korea.
     \label{Jeong}
   }
     \date{Received 12/4/2018; accepted 20/4/2019}

   
     \abstract{Observations of the innermost regions of deeply embedded protostellar cores have revealed complicated physical structures as well as a rich chemistry with the existence of complex organic molecules. The protostellar envelopes, outflow and large-scale chemistry of Class~0 and Class~I objects have been well-studied, but while previous works have hinted at or found a few Keplerian disks at the Class~0 stage, it remains to be seen if their presence in this early stage is the norm. Likewise, while complex organics have been detected toward some Class~0 objects, their distribution is unknown as they could reside in the hottest parts of the envelope, in the emerging disk itself or in other components of the protostellar system, such as shocked regions related to outflows.}
  {In this work, we aim to address two related issues regarding protostars: when rotationally supported disks form around deeply embedded protostars and where complex organic molecules reside in such objects. We wish to observe and constrain the velocity profile of the gas kinematics near the central protostar and determine whether Keplerian motion or an infalling-rotating collapse under angular momentum conservation best explains the observations. The distribution of the complex organic molecules  are used to investigate whether they are associated with the hot inner envelope or a possible Keplerian disk.}
  {We observed the deeply embedded protostar, L483, using Atacama Large Millimeter/submillimeter Array (ALMA) Band~7 data from Cycles~1 and 3 with a high angular resolution down to $\sim$~0.1$^{\prime\prime}$ (20~au) scales. We present new HCN $J=4$--$3$, HCO$^{+}$ $J=4$--$3$, CS $J=7$--$6$, and H$^{13}$CN $J=4$--$3$ observations, along with a range of transitions that can be attributed to complex organics, including lines of CH$_3$OH, CH$_3$OCHO, C$_2$H$_5$OH, NH$_2$CHO, and other species.} 
  {We find that the kinematics of CS~$J=7$--$6$ and H$^{13}$CN~$J=4$--$3$ are best fitted by the velocity profile from infall under conservation of angular momentum and not by a Keplerian profile. The only discernible velocity profile from the complex organics, belonging to CH$_3$OCHO, is consistent with the infall velocity profile derived from CS~$J=7$--$6$ and H$^{13}$CN~$J=4$--$3$. The spatial extents of the observed complex organics are consistent with an estimated ice sublimation radius of the envelope at $\sim$~50~au, suggesting that the complex organics exists in the hot corino of L483, where the molecules sublimate off the dust grain ice-mantles and are injected into the gas phase.}
{We find that L483 does not harbor a Keplerian disk down to at least $15$~au in radius. Instead, the innermost regions of L483 are undergoing a rotating collapse, with the complex organics existing in a hot corino with a radius of $\sim$~40--60~au. This result highlights that some Class~0 objects contain only very small disks, or none at all, with the complex organic chemistry taking place on scales inside the hot corino of the envelope, in a region larger than the emerging disk.}


               
\offprints{Steffen Kj{\ae}r Jacobsen, \email{skjacresearch@gmail.com}}
   \maketitle
%
%

\section{Introduction}
\label{sec:intro}

Low-mass stars like our Sun are formed from the gravitational collapse of a dense core within a cold molecular cloud. The inherent rotation of the cloud necessitates the presence of outflows and jets in the system to transport angular momentum away and let the central protostar grow in mass. A protostellar disk will emerge around the growing protostar, due to the conservation of angular momentum of material not lost from the system \citep{1984ApJ...286..529T}. Another angular momentum loss mechanism, strong magnetic braking,  however, can prevent a disk from being formed altogether \citep{2011ppcd.book.....G}.
If a disk-like structure is able to form, it will eventually become rotationally supported, resulting in a Keplerian disk. Also, in the innermost, warmest part of the envelope, a rich chemistry should take place, with the sublimation of icy dust grain mantles leading to the presence of complex organic molecules in the gas phase \citep[e.g.,][]{herbst09}. These molecules may end up becoming part of the assembling circumstellar disk and thus incorporated into eventual planetary systems. It is therefore interesting to investigate the link between the physical and chemical structures of inner envelopes and emerging circumstellar disks, a topic where the Atacama Large Millimeter/submillimeter Array (ALMA) with its high angular resolution and sensitivity is ideally suited to make significant contributions. This paper presents observations down to a radius of $10$~au of the Class~0 protostar in the isolated core Lynds~483, with the aim of studying its chemistry and using its kinematical structure to shed light on its physical structure on these scales.

Concerted efforts have been made over some time on observing the innermost regions of deeply embedded protostars in the Class~0 and Class~I stage \citep[e.g.,][]{1998ApJ...502..315H, 2000ApJ...529..477L, 2004A&A...416..603J,2007ApJ...659..479J,2009A&A...507..861J, 2011ApJS..195...21E}, revealing excess compact dust emission that could be early disk-like structures or rotationally supported disks. The early evolution and exact formation time of these earliest Keplerian disks are not well-established, however, due to the difficulty of disentangling cloud and disk emission in interferometric observations, the low number of known Class~0 Keplerian disks and the unknown number of Class~0 objects lacking a rotationally supported disk. With the advent of high angular resolution interferometers such as ALMA, discovering early, relatively small disks has become feasible.
Keplerian disks are observed around Class~I objects on $\sim$ $100$~au scales \mbox{\citep{2007A&A...475..915B, 2009A&A...507..861J, 2014A&A...562A..77H}}  as well as around some Class~0 objects: for example, the Class~$0$/I protostar L1527 is found to have a Keplerian disk with a radius of $50$--$90$~au \mbox{\citep{2012Natur.492...83T, 2014ApJ...796..131O}}. Also, \cite{2013A&A...560A.103M} detect a disk around the Class~0 protostar VLA1623 that is rotationally supported with a Keplerian profile out to at least $150$~au, \cite{2014A&A...566A..74L} report a $50$~au Keplerian disk around the Class~$0$/I protostar R~CrA-IR$7$B, and \citet{2014A&A...568L...5C} make a tentative detection of a $90$~au Keplerian disk around the Class~0 protostar HH212. On the other hand, the Class~0 object B$335$ is shown to lack an observable Keplerian disk down to a radius of $10$~au \citep{2015ApJ...812..129Y} and continuum emission in the innermost region of B$335$ is consistent with only a very small disk mass \citep{2015ApJ...814...22E}.  Due to the uncertain nature of some of these Keplerian disk detections and the small sample size, more detections of rotationally supported disks in the earliest stages, or equivalently, non-detections and upper  limits to the sizes of disks around Class~0 objects, are needed to constrain disk formation theories.

Concurrently with the investigation of Class~0 and I disks, hot regions in the innermost parts of envelopes hosting low-mass star formation have been observed. Such regions have been linked to the formation of Complex Organic Molecules (COMs). Called a 'hot corino' in the case of a low-mass star, these regions of hot gas, T~>~$90$--$100$~K, are where the icy dust mantles composed of different molecules sublimate and the molecules are released into the gas phase, in which COMs and prebiotic molecules have been discovered \citep[e.g.,][]{2004ApJ...617L..69B,2005ApJ...632..973J, 2012ApJ...757L...4J,2015A&A...576A...5C, 2015ApJ...804...81T}.  The presence of COMs has also been linked to the transition region between the outer infalling-rotating envelope and the centrifugal barrier, i.e., the radius where the kinetic energy of the infalling material is converted into rotational energy \citep{2014Natur.507...78S}. Accretion shocks and other heating events in this transition zone are hypothesized to induce a chemical change \citep[e.g.,][]{2014Natur.507...78S, 2016ApJ...824...88O}. From an astrochemical point of view, mapping the molecular inventory and distribution at this early stage of the disk, or even before the disk is formed, will set the stage for subsequent chemical evolution, all the way up to the more complex, prebiotic molecules. 

An interesting object for addressing these issues is the dense core, Lynds 483 (L483), constituting the envelope around the Class~0 infrared source IRAS~18148-0440\footnote{In the literature both the core and the infrared source are referred to as L483, which we follow for the remainder of the paper}.  Traditionally, L483 has been associated with the Aquila Rifts region at a distance of 200~pc \citep{1985ApJ...297..751D}. Recently the distance to Aquila has been revised upward to 436$\pm$9~pc based on VLBA and Gaia-DR2 astrometry \citep{ortizleon18}. Yet, Gaia-DR2 measurements of parallaxes and extinction of stars localised around L483, still suggest that the core itself is located at a closer distance of 200--250~pc (Appendix~\ref{distance}) and therefore that it is isolated and not physically associated with the larger scale cloud environment of Serpens/Aquila. We therefore adopt the previous distance estimate of 200~pc in this paper.  At this distance, the bolometric luminosity of L483 is $10$--$14$~L$_{\odot}$ \citep{1991ApJ...366..203L, 2000A&A...359..967T} making it one of the more luminous solar-type protostars and a good target for chemical studies.

L483 drives a well-collimated bipolar CO outflow \citep{1988MNRAS.235..139P, 1991MNRAS.252..442P,1995ApJ...453..754F,1996A&A...311..858B,1999A&A...344..687H}. Also, it is associated with a variable H$_2$O maser \citep{1995ApJS...99..121X} and shocked H$_2$ emission, which is suggested to originate from the head and edges of the jet where it interacts with ambient molecular gas \citep{1995ApJ...453..754F}. \citet{2013ApJ...770..151C} find the  position angle of a suggested magnetic pseudodisk to be 36$^{\circ}$ based on 4.5~$\upmu$m \textit{Spitzer} imaging, while the outflow  position angle is estimated at $105$$^{\circ}$, based on the shocked H$_2$ emission. \citet{1995ApJ...453..754F} found the outflow inclination of L483 to be 40$^{\circ}$ based on 2.22~$\upmu$m imaging, while analysis by \cite{2018ApJ...863...72O} of the CS and CCH line emission associated with the outflows found the outflow inclination angle to be between 75$^{\circ}$ and 90$^{\circ}$, that is, nearly perpendicular to the line-of-sight. In terms of its spectral energy distribution (SED) and envelope mass \citep[4.4~M$_\odot$;][]{2004A&A...424..589J} L483 appears as a deeply embedded Class~0 protostar. However, \citet{2000A&A...359..967T} find that its bipolar outflow has characteristics seen in both Class~0 and Class~I objects, and therefore propose that L483 is in transition from Class~0 to Class~I.

\cite{2000ApJS..131..249S} made 450~$\upmu$m and 850~$\upmu$m continuum maps of L483 with SCUBA at the JCMT, revealing its elongated continuum emission in the outflow direction, likely the outer parts of the envelope being swept up by the outflowing material. \citet{2004A&A...424..589J} find that the velocity gradients in HCN, CS, and N$_2$H$^+$ around the source are perpendicular to its outflows, indicative of a large-scale, infalling-rotating envelope, with the velocity vector being consistent with rotation around a central object of $\sim$1~M$_{\odot}$.  Curiously, the interferometric flux of L483 is consistent with envelope-only emission and does not require a central compact emission source \mbox{\citep{2007ApJ...659..479J, 2009A&A...507..861J, 2004A&A...424..589J}}. \cite{2017ApJ...837..174O} use a rotating collapse ballistic model and find that a 0.1--0.2 M$_{\odot}$ central protostar with a collapsing-rotating envelope with a centrifugal barrier radius (where the barrier radius is half of the centrifugal radius) of 30--200 au, assuming an inclination angle of 80$^{\circ}$, can roughly explain the observed CS, SO, HNCO, NH$_2$CHO, and HCOOCH$_3$ lines. \citet{2017ApJ...837..174O} suggest that some molecular species observed towards L483 may be in a Keplerian disk, very near the protostar, but it remains unclear whether the COMs are more directly linked to a Keplerian disk or to a hot corino region, which may not contain a Keplerian disk.

In this work, we use high angular resolution from ALMA Cycles~1 and 3 to image the distribution of COMs as well as to probe the kinematics of the innermost regions down to a radius of $\sim10$~au. These data enable us to improve our understanding of disk formation and the early astrochemistry of low-mass protostars.  This paper is structured as follows: First, the observations are described in Section~\ref{sec:obs}, while the results are presented in Section~\ref{sec:results}. An analysis of the inner region kinematics is presented, first for H$^{13}$CN~$J=4$--$3$ and CS~$J=7$--$6$, in Section~\ref{sec:kin}, and then for the observed COMs, together with an analysis of the dust temperature profile of the innermost region using a simple dust density model, in Section~\ref{sec:kin_coms}. Our results and analysis are discussed in Section \ref{sec:disc} and the conclusions in Section \ref{sec:conc}.

\section{Observations}
\label{sec:obs}
L483 was observed on the nights of $2013$ June $1$, $2013$ June $19$, $2013$ November $2$, and $2013$ November $3$ in ALMA Band~$7$ as part of Cycle~$1$ observations (PI. N.~J. Evans II, projectid: $2012.1.00346$.S) and on $2016$ August $31$, $2016$ September $7$, and $2016$ September $9$ in ALMA Band~$7$ as part of ALMA Cycle~$3$ observations (PI: J.~K. J\o rgensen, projectid: $2015.1.00377$.S).
Both observations were centered on $\alpha_{2000}$=18$^{\mathrm{h}}$17$^{\mathrm{m}}$29.90$^{\mathrm{s}}$, $\delta_{2000}$=-04$^{\circ}$39$^{\prime}$39.50$^{\prime\prime}$, with total integration times of 1.8~hours and 3.6~hours for Cycle~1 and Cycle~3, respectively. The Cycle~1 observations used 44 12-m antennas, with baselines in the range of $20$--$600$~k$\lambda$, while the Cycle~3 observations used either $38$ or $39$ $12$-m antennas with baselines in the range of $15$--$1800$~k$\lambda$.
\begin{table*}
\centering
\caption{Observational spectral windows.}
\begin{tabular}{l*{8}{c}r}\toprule
\midrule
ID & Frequency range [GHz] & rms [mJy] & Channelwidth [MHz] & Synthesized beam \\ 
\midrule
\textit{Combined Cycle~1 and 3 data} \\
\hline
0 & 342.766 -- 343.000 & 6 & 0.244 & 0.36$^{\prime\prime}\times0.26^{\prime\prime}$ \\  
1 & $345.215$ -- 345.450 & 5 & 0.244 & 0.28$^{\prime\prime}\times0.21^{\prime\prime}$ \\ 
2 & 354.379 -- 354.613 & 6 & 0.244 & 0.36$^{\prime\prime}\times0.26^{\prime\prime}$ \\  
3 & 356.611 -- 356.845 & 7 & 0.244 & 0.30$^{\prime\prime}\times0.20^{\prime\prime}$ \\ 
\hline
\textit{Cycle~3} \\
\hline
0 & 342.757 -- 343.000 & 4 & 0.244 & 0.13$^{\prime\prime}\times0.11^{\prime\prime}$ \\
1 & 345.097 -- 345.565 & 3 & 0.244 & 0.14$^{\prime\prime}\times0.13^{\prime\prime}$ \\
2 & 354.261 -- 354.730 & 4 & 0.244 & 0.13$^{\prime\prime}\times0.13^{\prime\prime}$ \\
3 & 356.490 -- 356.958 & 4 & 0.244 & 0.13$^{\prime\prime}\times0.13^{\prime\prime}$ \\

\bottomrule
\end{tabular}
\label{tab:spect_windows_cyc13}
\caption*{Frequency corresponds to $v_{\mathrm{lsr}}$~=~0~km~s$^{-1}$. The rms is given as the typical rms in flux~beam$^{-1}$ in the channels.}
\end{table*}

For Cycle~1, L$483$ was observed on $2013$ June $01$ with J$1733$--$1304$ as the phase and flux calibrator and J$1924$-$2914$ as the bandpass calibrator. For $2013$ June $19$, J$1733$-$1304$ was the phase, flux, and bandpass calibrator. For $2013$ November $2$ and $3$, J$1733$--$1304$ was the phase calibrator and J$1924$--$2914$ was the flux and bandpass calibrator. 
For Cycle~3, L$483$ was observed on 2016 August 31 with J$1924$-$2914$ as the bandpass and flux calibrator, and J$1743$-$0350$ as the phase calibrator. For $2016$ September $7$ and $9$, J1751+0939 was the bandpass calibrator, while J1733-1304 was the flux calibrator, and J1743-0350 was the phase calibrator.  
Both Cycle~1 and 3 data were calibrated using \texttt{CASA} v. 4.7. Before combination of the datasets, the Cycle~1 data were binned down with 4~channels in each bin to match the Cycle~3 channel width, as the Cycle~1 observations had higher spectral resolution than those of Cycle~3. Also, the Cycle~3 data were trimmed at the spectral window edges, to match the Cycle~1 bandwidth (see Table~\ref{tab:spect_windows_cyc13} for spectral window details).

Phase self-calibration was also performed on the continuum channels in each dataset before combination.
After concatenation of the Cycles~1 and 3 data, the continuum was constructed using line-free channels and subtracted from the line emission cubes. After primary beam correction, both line emission channel images and the 857~$\upmu$m continuum image were created with the \texttt{clean} algorithm using Briggs weighting with a robust parameter of 0.5, to get a good trade-off between sensitivity and angular resolution. 

An 857~$\upmu$m continuum image and line emission cubes were also constructed using phase self-calibrated Cycle~3 data alone, to investigate the spectrum of the broader bandwidth in the Cycle~3 observations, and to investigate the spatial distribution of COMs on the smallest scales (Table~\ref{tab:spect_windows_cyc13}).

\begin{figure}
 \includegraphics[scale=0.65]{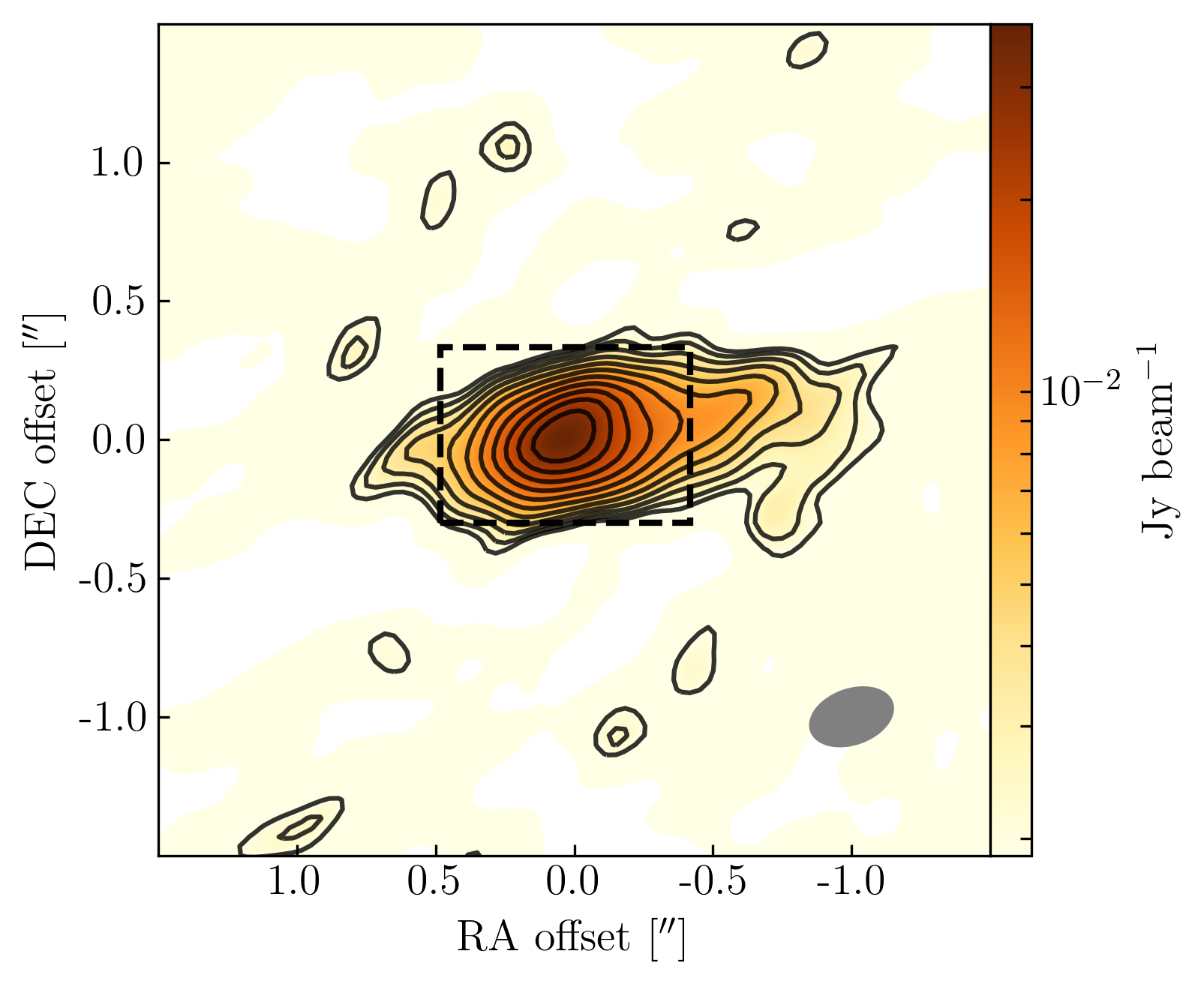}
 \caption{Cycles~1 and 3 combined dataset continuum emission at 857~$\upmu$m in logscale. The coordinates are centered on the continuum center at $\alpha_{2000}$=18$^{\mathrm{h}}$17$^{\mathrm{m}}$29.942$^{\mathrm{s}}$, $\delta_{2000}$=-04$^{\circ}$39$^{\prime}$39.597$^{\prime\prime}$. Contours are spaced logarithmically between $5$--$100$~\% of the peak emission, in 12 steps. The box marks the region used for the continuum analysis in Section \ref{sec:kin_coms}.}
 \label{fig:dust_cont}
\end{figure}
\begin{figure*}
 \includegraphics[width=\textwidth]{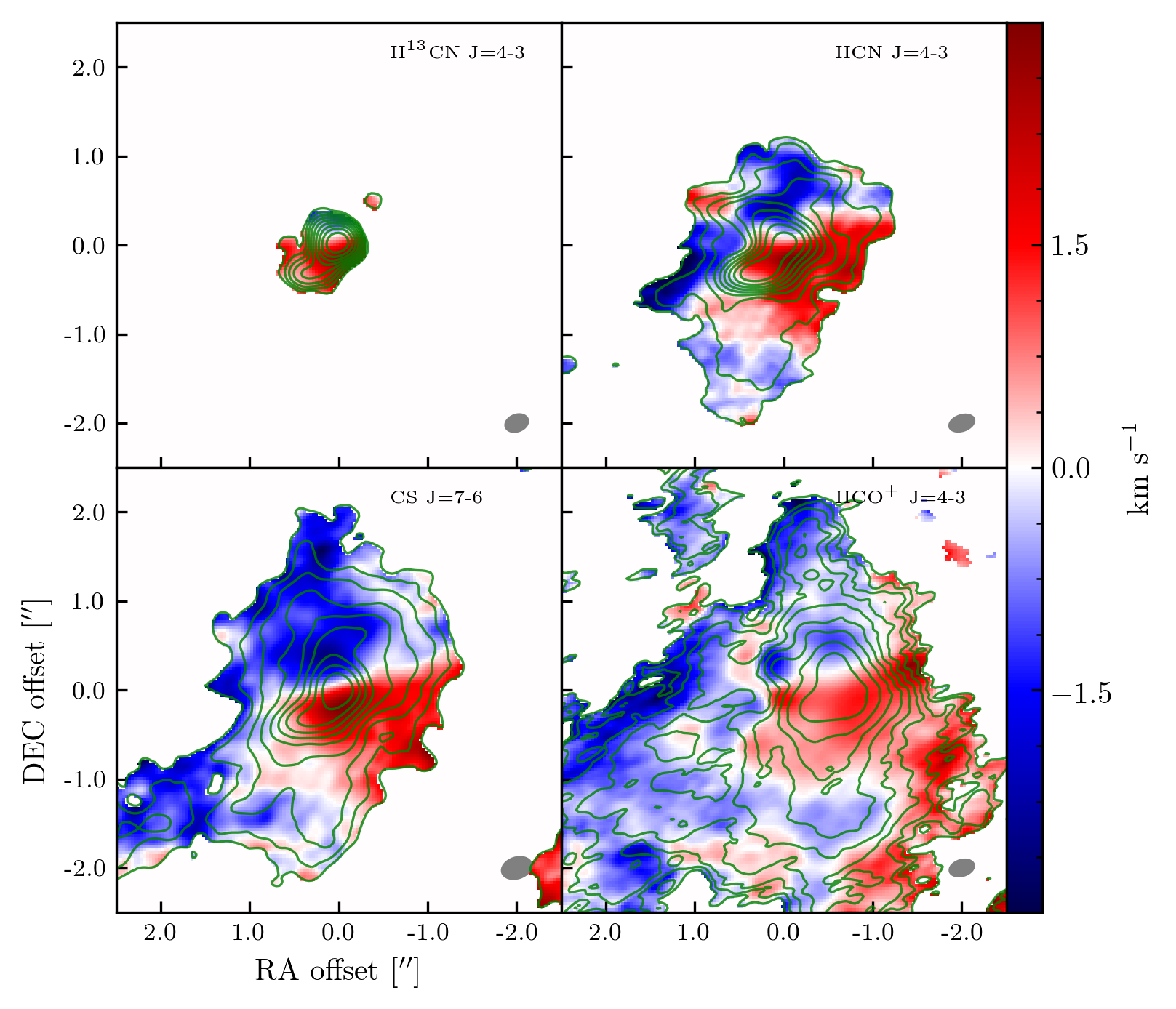}
 \caption{Moment~$0$ and $1$ maps of H$^{13}$CN~$J$~=~4--3, CS~$J$~=~7--6, HCN~$J$~=~4--3, and HCO$^{+}$~$J$~=~4--3. Moment~$0$ maps are in green contours, overlaid on the moment~1 maps. The moment~$0$ map contours are spaced logarithmically between $5$--$100$~\% of the peak emission, in 10 steps. The peak emissions are 0.94, 1.2, 1.2, and 0.69 Jy~Beam$^{-1}$ km s$^{-1}$ for H$^{13}$CN~$J$~=~4--3, CS~$J$~=~7--6, HCN~$J$~=~4--3, and HCO$^{+}$~$J$~=~4--3, respectively. The mean rms is 0.22~Jy~Beam$^{-1}$ km~s$^{-1}$. All the data shown in this figure are from the combined dataset.}
 \label{fig:mom_maps}
\end{figure*}

\section{Results}
\label{sec:results}
Fig.~\ref{fig:dust_cont} shows the combined dust continuum image at 857~$\upmu$m, revealing concentrated emission with a deconvolved 2D Gaussian fit of 0.36$^{\prime\prime}\times$0.20$^{\prime\prime}$ at a position angle of 104$^{\circ}$. The dust is elongated in the East-West direction, possibly caused by the outflows dragging material with it outwards.
We estimate the integrated deconvolved dust continuum of the combined Cycle 1+3 dataset at $857$~$\upmu$m (Fig.~\ref{fig:dust_cont}) to be 68.6 $\pm$ 4.4~mJy,  while the Cycle~3-only continuum image has a deconvolved size of $0.16^{\prime\prime}\times0.13^{\prime\prime}$, with a position angle of $156^{\circ}$ and an integrated flux density of $53.7$~$\pm$~$3.5$ mJy. These value are fairly close to the SMA $0.8$~mm integrated flux density \mbox{\citep{2007ApJ...659..479J}} of $97$~mJy (point source fit with baselines longer than $40$~k$\lambda$), but lower than a circular fit to the SMA continuum visibilities (integrated flux of 0.2 Jy)  an indication that the ALMA observations filter out some of the larger-scale emission. 

Fig.~\ref{fig:mom_maps} shows emission from the four major molecule transitions in the observed spectral windows, H$^{13}$CN~$J$~=~4--3, HCN~$J$~=~4--3, CS~$J$~=~7--6, and HCO$^{+}$~$J$~=~4--3. Beyond compact emission coincident with the continuum source, each shows an extra emission component South-East of the disk itself (Fig.~\ref{fig:mom_maps}), which we interpret as the inner domain of the eastward part of the known bipolar outflow. A velocity gradient in the North-South direction is also observed in the four major emission lines. 

\begin{figure*}
 \includegraphics[width=\textwidth]{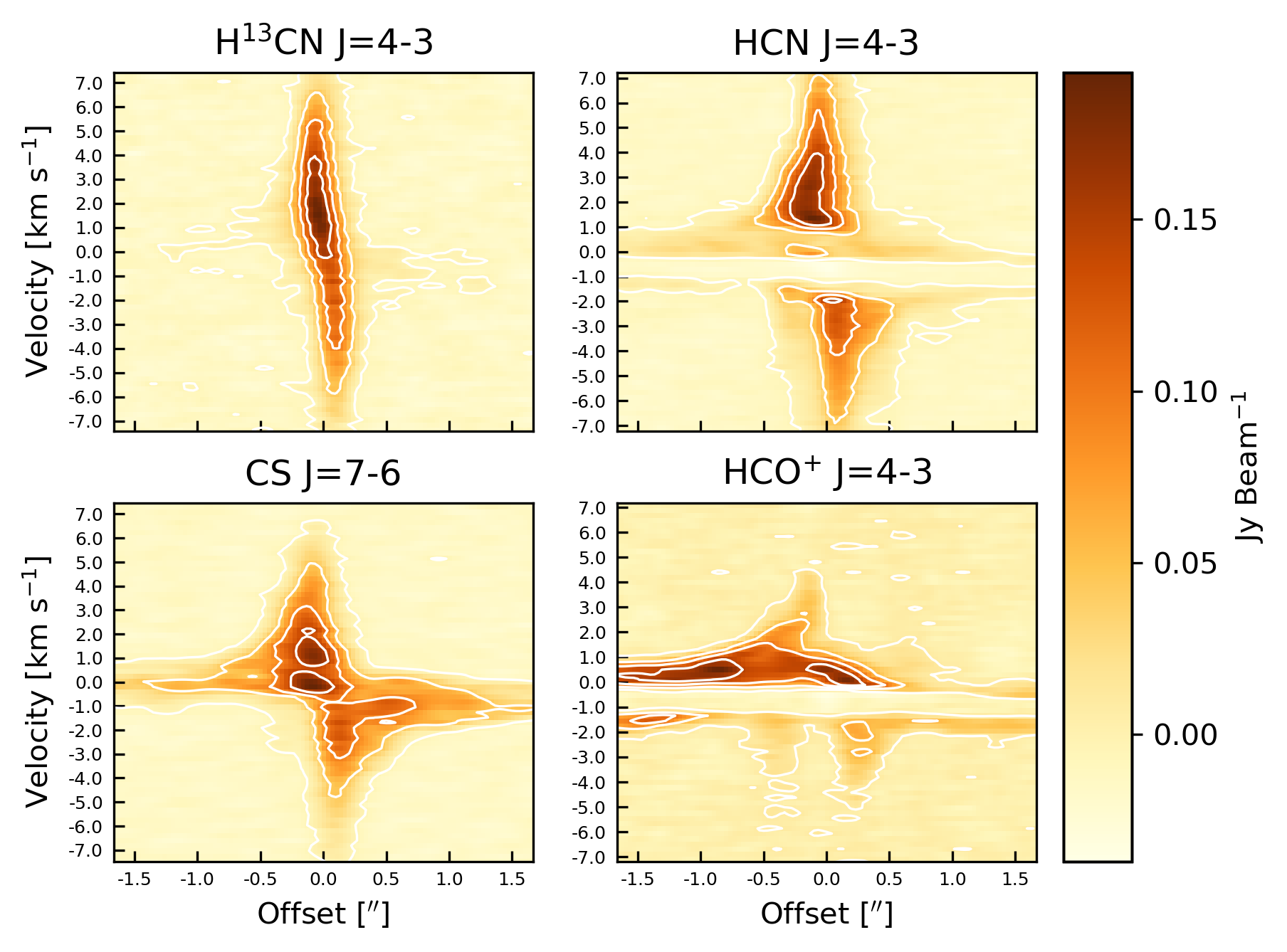}
 \caption{PV diagrams of H$^{13}$CN~$J$~=~4--3, CS~$J$~=~7--6, HCN~$J$~=~4--3, and HCO$^{+}$~$J$~=~4--3, using $v_{\mathrm{lsr}}=6.0$~km~s$^{-1}$. Contours are spaced linearly between $5$--$100$~\% of the peak emission, in five steps. The PV diagrams are made along the direction of the velocity vector and the offset is the distance to the rotation axis (Fig.~\ref{fig:mom_cont_h13cn}). The emission center, as defined in Section~\ref{sec:anal}, is the intersection of the rotation axis and velocity vector. $v_{\mathrm{lsr}}=6.0$~km~s$^{-1}$, derived in Section~\ref{sec:anal}, matches the H$^{13}$CN~$J$~=~4--3 emission well, while the large-scale emission seen in \HCN, \CS, and \HCO are better matched by 5.5~km$^{-1}$ \mbox{\citep{1999A&A...344..687H}}. All the data shown in this figure are from the combined dataset.}
 \label{fig:pv}
\end{figure*}

\begin{figure*}
\includegraphics[scale=1.2]{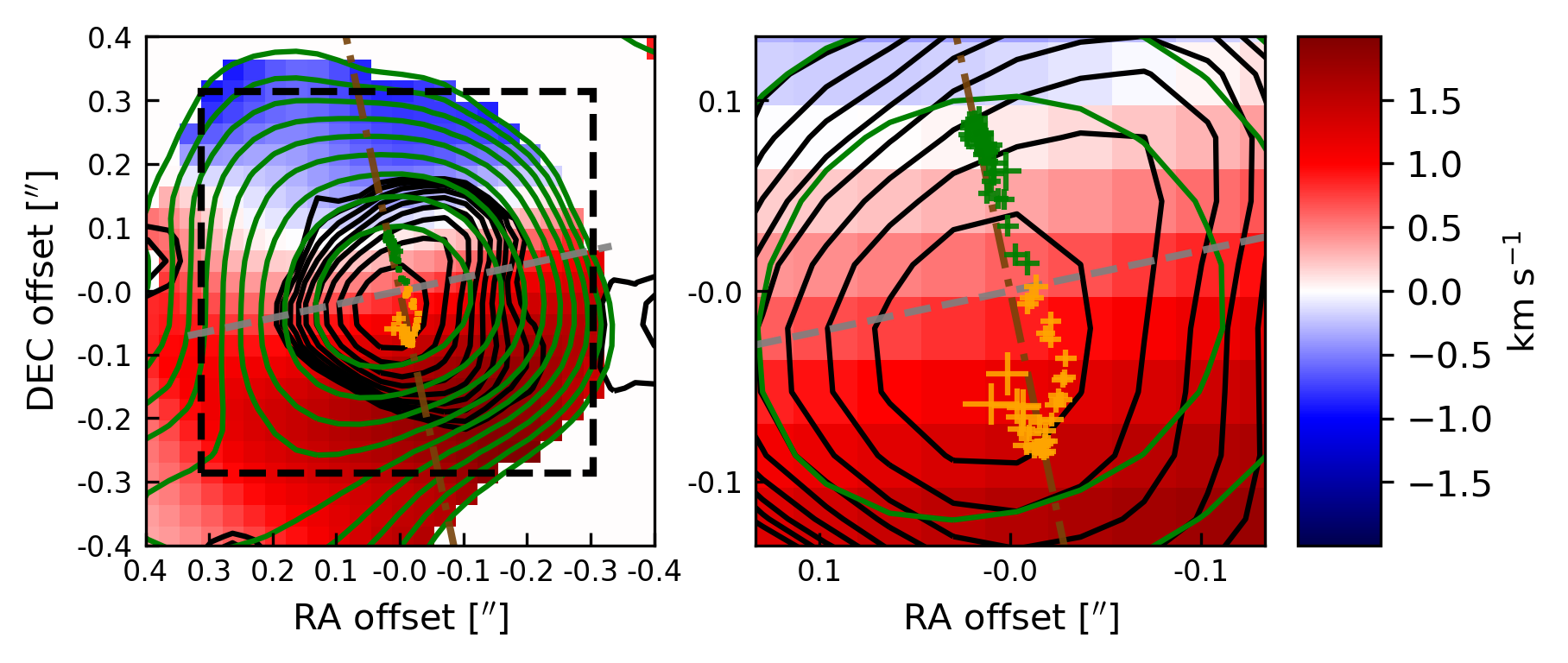}
\caption{H$^{13}$CN~$J$~=~$4$--$3$ \texttt{imfit} datapoints superimposed on the H$^{13}$CN~$J$~=~4--3 moment~1 map, green and orange points indicating blue- and redshifted channels, respectively. The width and height of the data symbols indicate the 1$\sigma$ uncertainty of the peak position calculated by \texttt{imfit}. Green and black contours show the H$^{13}$CN~$J$~=~4--3 moment~0 map and 857~$\upmu$m continuum emission, respectively, both spaced logarithmically between $5$--$100$~\% of the peak emission, in 10 steps. The inferred rotation axis of H$^{13}$CN~$J$~=~4--3 is shown as a gray dashed line and the velocity gradient vector as a brown dashed line. The emission center is the intersection between the rotation axis and the velocity vector. The black dashed box indicate the region used for the \texttt{imfit} routine and also shows the region from which the spectra in Fig.~\ref{fig:spect_windows} were extracted. All the data shown in this figure are from the Cycle~3 dataset.} 
 \label{fig:mom_cont_h13cn}
\end{figure*}

\begin{figure*}
\includegraphics[scale=1.2]{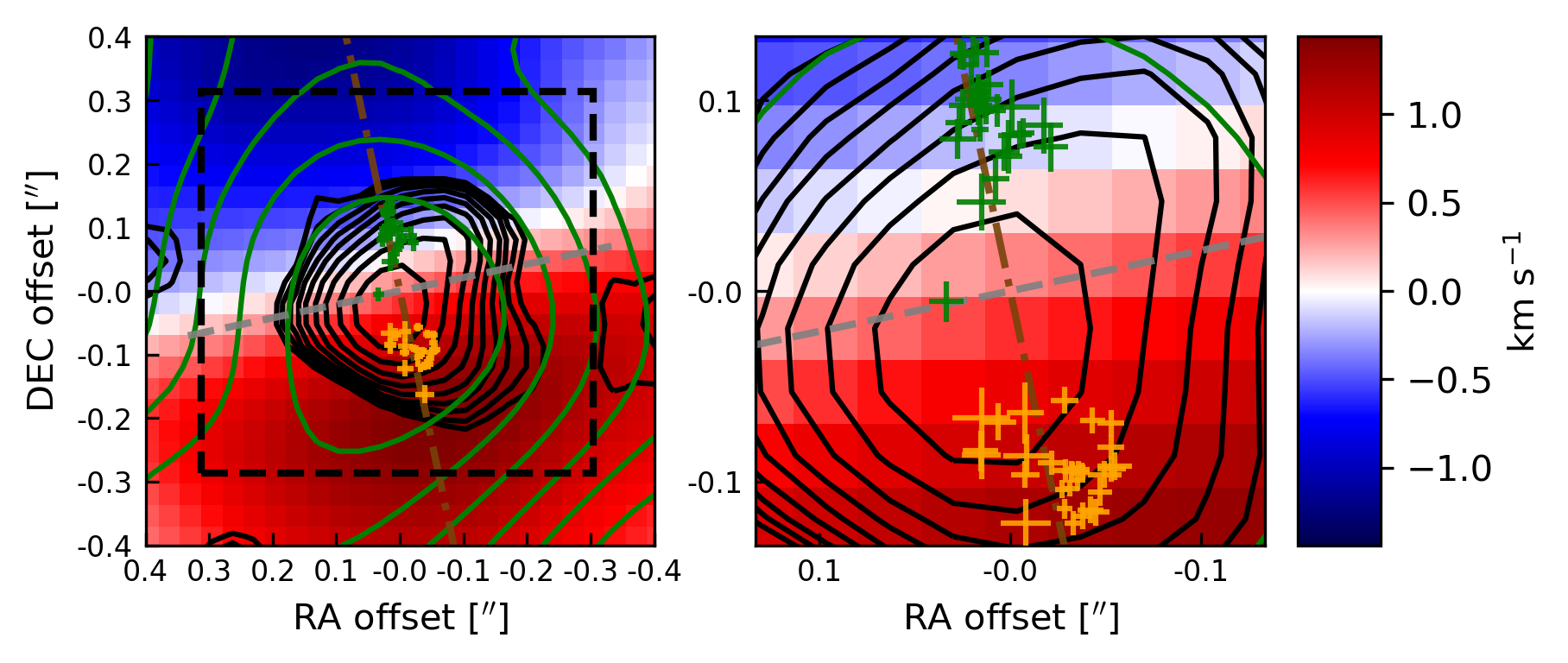}
\caption{CS~$J$~=~$7$--$6$ \texttt{imfit} datapoints superimposed on the CS~$J$~=~7--6 moment~1 map. Green and orange points indicates blue- and redshifted channels, respectively. The width and height of the data symbols indicate the 1$\sigma$ uncertainty of the peak position calculated by \texttt{imfit}. Green and black contours show the CS~$J$~=~7--6 moment~0 map and 857~$\upmu$m continuum emission, respectively, both spaced logarithmically between $5$--$100$~\% of the peak emission, in 10 steps. The velocity vector (brown) and rotation axis (grey) was not fitted to the CS~$J$~=~7--6 data, but to H$^{13}$CN~$J$~=~4--3 instead, in order for the data from the two molecule transitions to be compared in the distance-velocity plot in Fig.~\ref{fig:imfit}. The black dashed box indicates the region used for the \texttt{imfit} routine and also shows the region from which the spectra in Fig.~\ref{fig:spect_windows} were extracted. All the data shown in this figure are from the Cycle~3 dataset.} 
 \label{fig:mom_cont_cs}
\end{figure*}

Fig.~\ref{fig:pv} shows PV diagrams of H$^{13}$CN~$J$~=~4--3, HCN~$J$~=~4--3, CS~$J$~=~7--6, and HCO$^{+}$~$J$~=~4--3. Each was produced along the velocity vector using \texttt{PVEXTRACTOR}, with a path width of 0.05$^{\prime\prime}$, where the offset is defined by the distance to the rotation axis (Fig.~\ref{fig:mom_cont_h13cn} and Fig.~\ref{fig:mom_cont_cs}). The PV diagrams show that only H$^{13}$CN~$J$~=~4--3 lacks extended structures and that HCN~$J$~=~4--3 and HCO$^{+}$~$J$~=~4--3 are heavily absorbed near the rest velocity, which is also seen in the observations of the entire envelope. All molecule transitions show asymmetric lines, with redshifted emission of higher intensity than the blueshifted emission. 

\mbox{\citet{2017ApJ...837..174O}} observed a compact component with a broad velocity width of $v$~=~$\pm$~$6$~km~s$^{-1}$ in CS~$J$~=~5--4, which we observe as well in our CS~$J$~=~7--6 emission.
While absorption against the continuum is observed near the rest velocity in \HCO and \HCN (Fig.~\ref{fig:spect_windows}), it is not significantly redshifted as expected for infalling material. The absorption we do see is likely caused by resolving out large-scale emission, due to a lack of shorter baselines. While redshifted absorption is seen towards similar Class~0 objects, such as B$335$ \mbox{\citep{2015ApJ...814...22E}}, L483 has an elongated structure, with outflows at an inclination of 75--90$^{\circ}$ and the kinematics of the innermost region, where we have extracted our spectrum, are also dominated by rotational motion, not free-fall (Fig.~\ref{fig:mom_maps}). This complex geometry and the kinematics may explain why the inverse P Cygni profile with redshifted absorption against the continuum, expected for a spherical collapse, is not present in L483.  

\subsection{Other lines and their identification}
\begin{figure*} 
  \begin{subfigure}[b]{0.5\textwidth}
    \includegraphics[width=\textwidth]{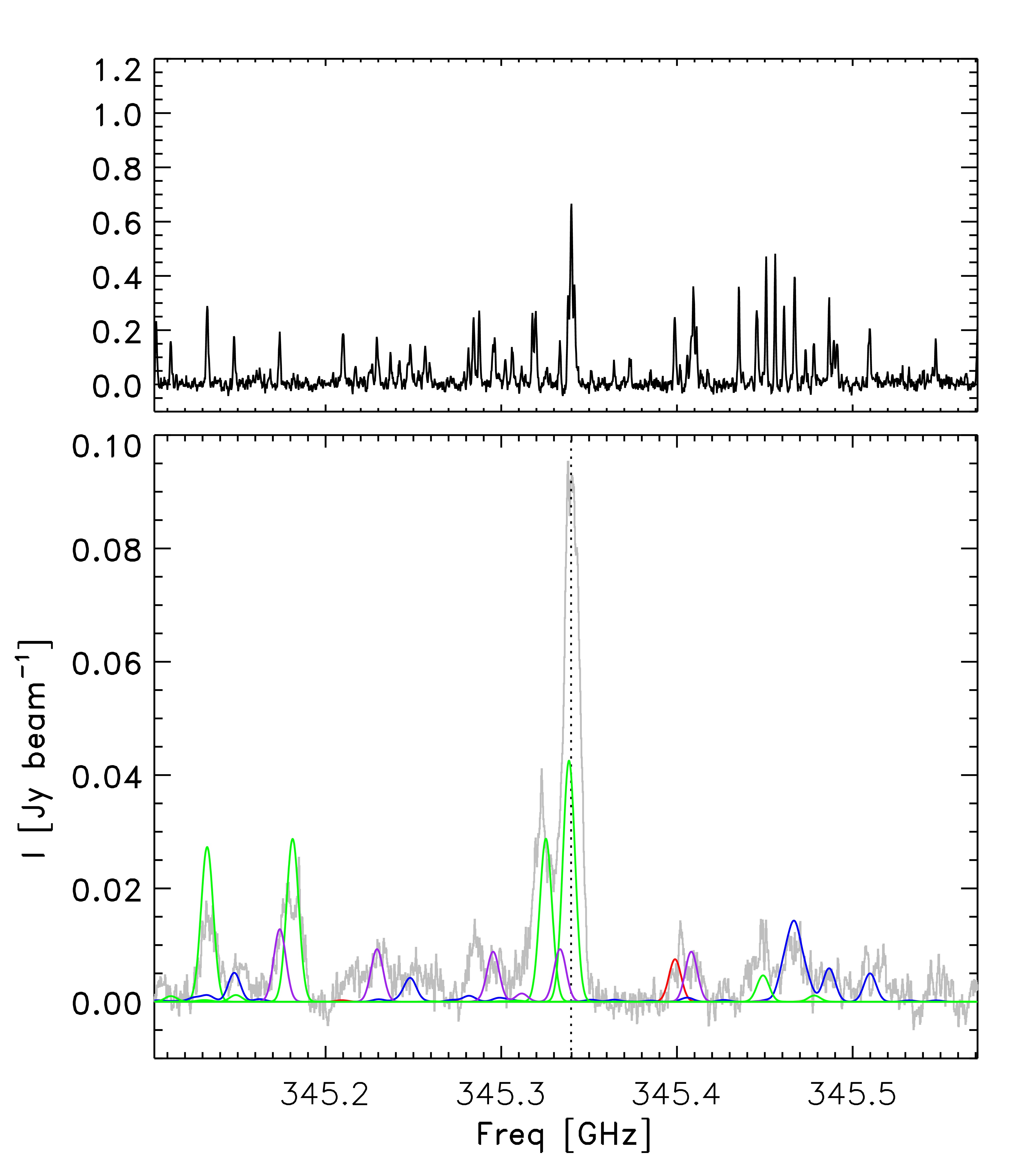}
    \caption{Spectral window centered on H$^{13}$CN~$J$~=~4--3.} 
    \label{fig:h13cn_spectrum} 
  \end{subfigure} 
  \begin{subfigure}[b]{0.5\textwidth} 
    \includegraphics[width=\textwidth]{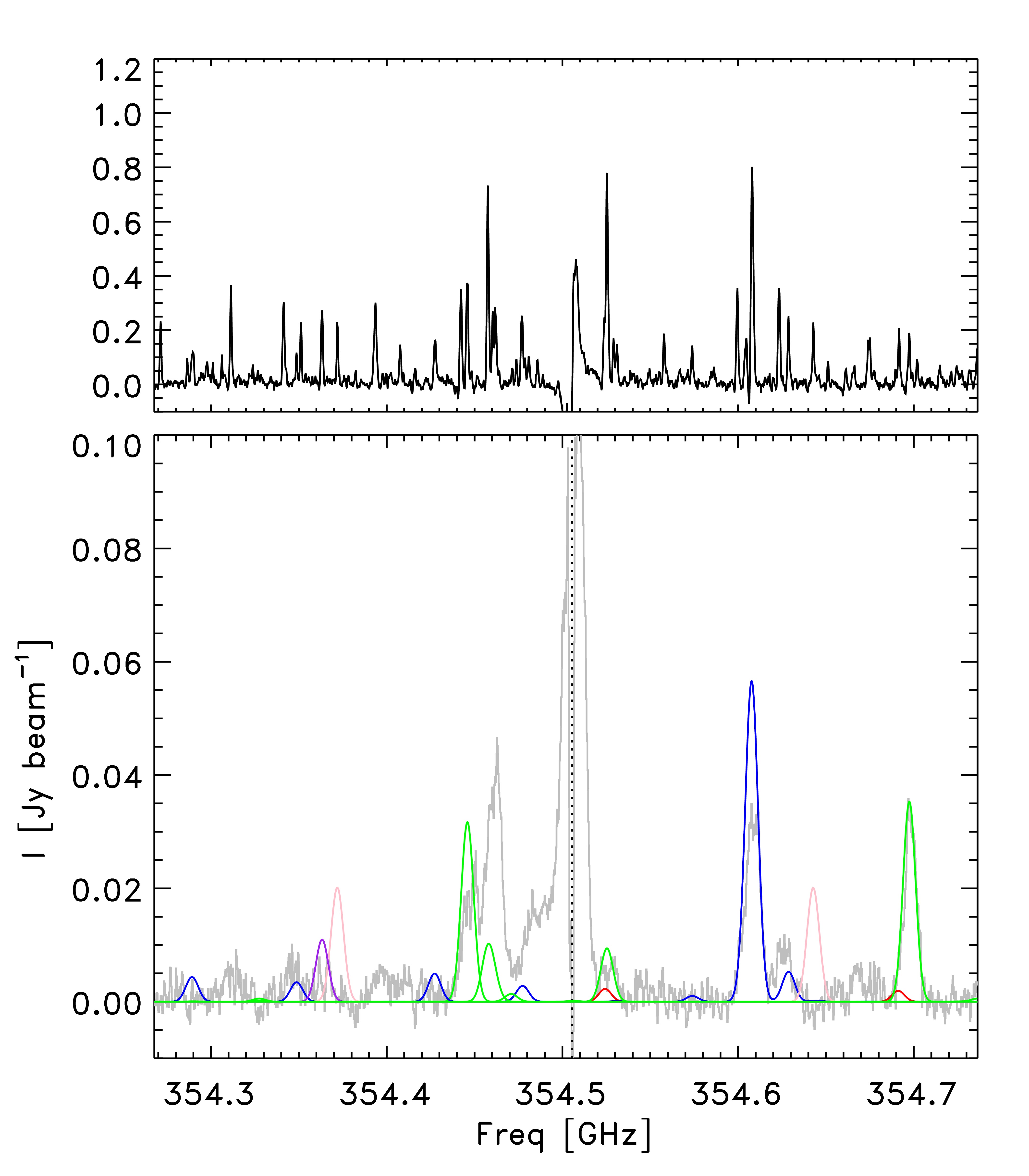} 
    \caption{Spectral window centered on HCN~$J$~=~4--3.}  
    \label{fig:hcn_spectrum} 
  \end{subfigure} 
  \begin{subfigure}[b]{0.5\textwidth} 
    \includegraphics[width=\textwidth]{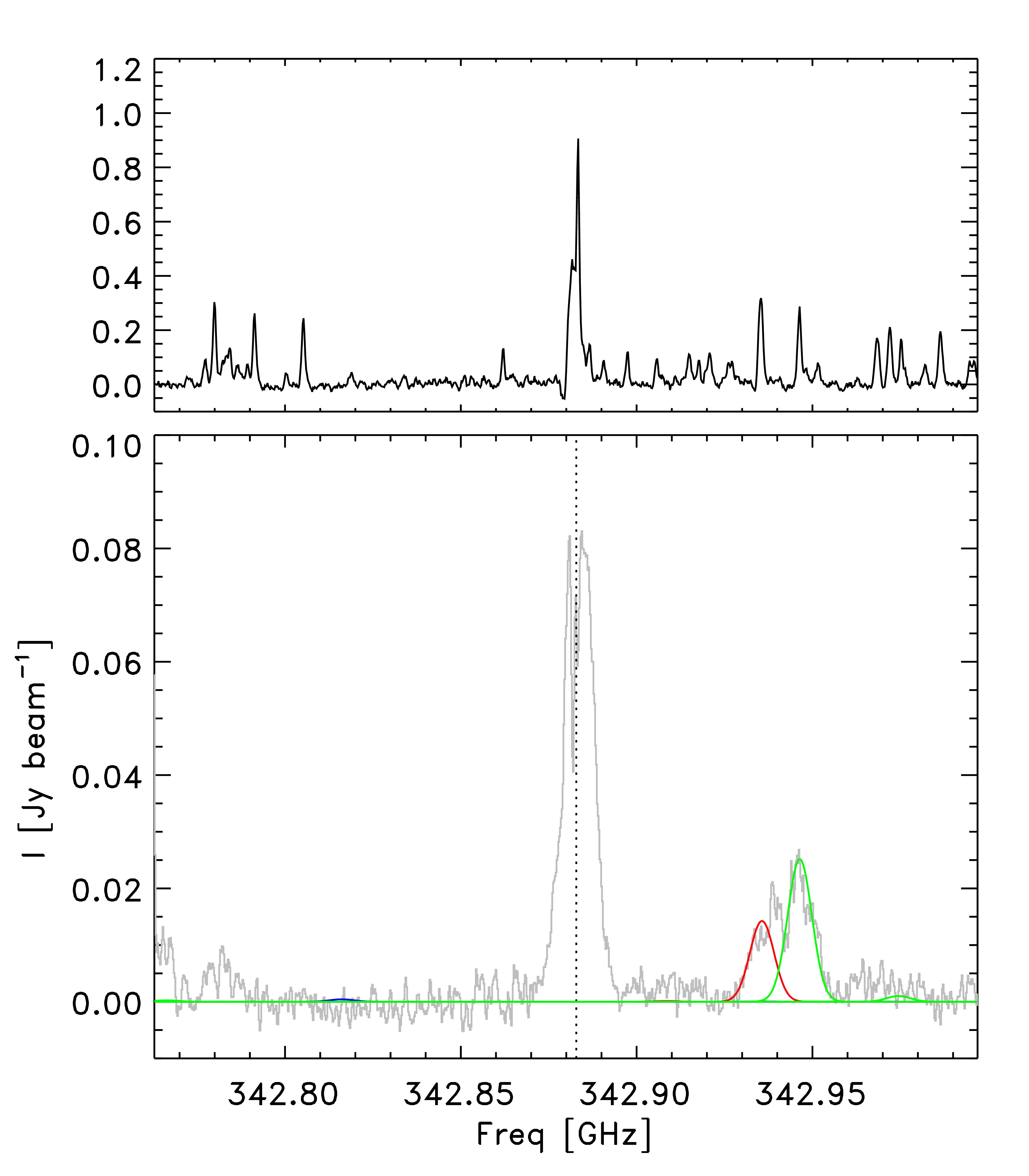} 
    \caption{Spectral window centered on CS~$J$=~7--6.}  
    \label{fig:cs_spectrum} 
  \end{subfigure}
  \begin{subfigure}[b]{0.5\textwidth} 
    \includegraphics[width=\textwidth]{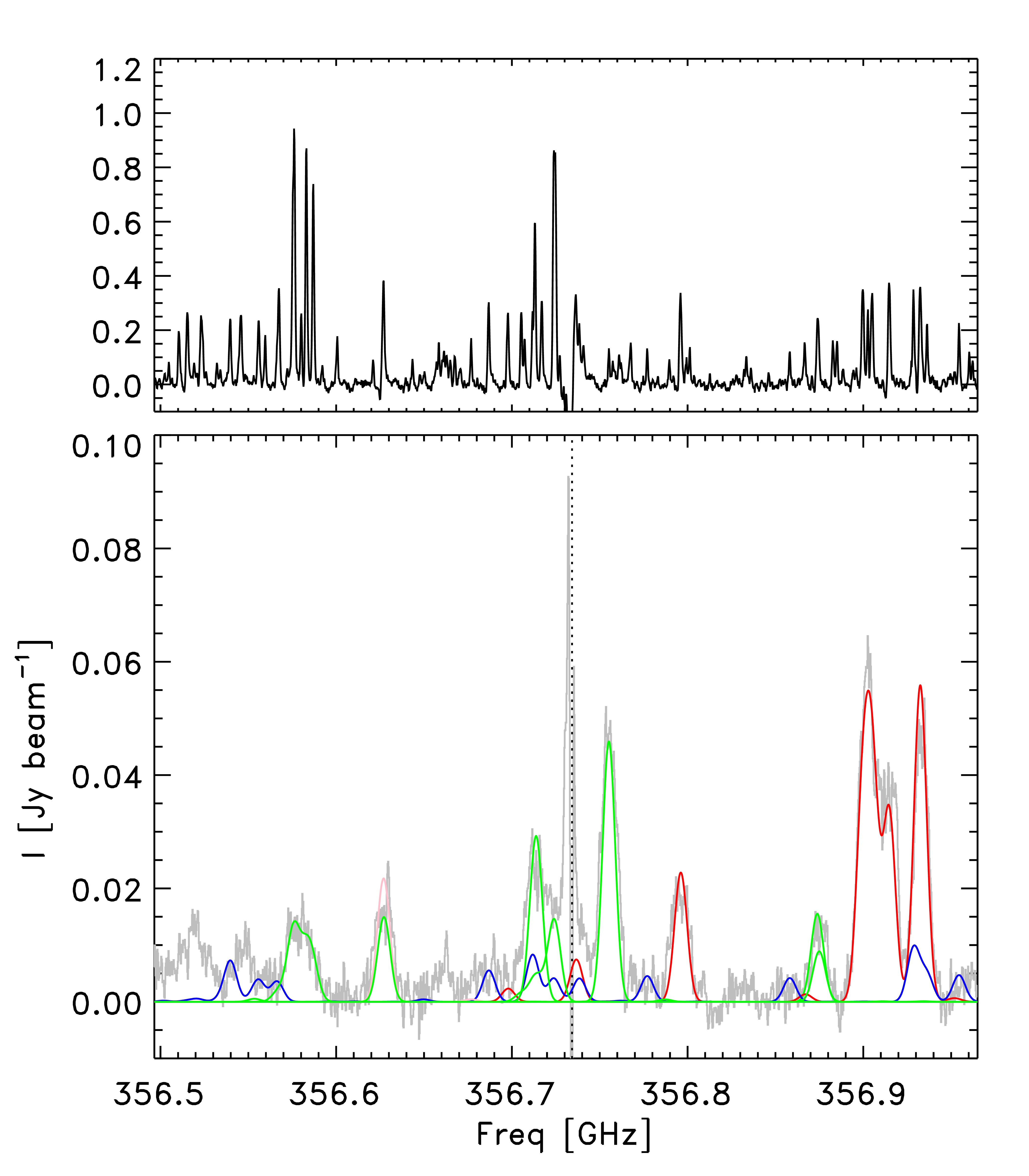} 
    \caption{Spectral window centered on HCO$^{+}$~$J$~=~4--3.}
    \label{fig:hco_spectrum}
  \end{subfigure}
  \caption{Cycle~3 spectral windows centered on H$^{13}$CN~$J$~=~$4$--$3$, HCN~$J$~=~4--3, CS~$J$~=~7--6, and HCO$^{+}$~$J$~=~4--3. The main transition of the spectral window is shown as a dotted line, assuming $v_\mathrm{lsr}$~=~$5.4$~km~s$^{-1}$. The line model fits are overlaid in colored lines (red~=~CH$_2$DOH, blue~=~CH$_3$OCHO, purple~=~C$_2$H$_5$OH, pink~=~CH$_3$SH, green~=~all other species), while the observed spectrum is in gray. The top frames show the corresponding IRAS16293B spectrum. The region from which the spectra was extracted can be seen in Fig.~\ref{fig:mom_cont_h13cn} and \ref{fig:mom_cont_cs}.}
  \label{fig:spect_windows}
\end{figure*}

In addition to the four main species (HCN, H$^{13}$CN, CS, and HCO$^+$) targeted as part of this program, a multitude of emission lines are found, belonging to a range of COMs, shown in Fig.~\ref{fig:spect_windows} along with their synthetic line spectra. We investigated COMs in the Cycle~3 spectra only, due to its larger frequency range. Identifying these fainter transitions in the spectra requires careful
comparison to spectroscopic catalogs, modeling the emission from known
species, and comparison to other surveys. To do this we calculate synthetic spectra for possible molecules and compared those to the data: for a given molecule the spectra are predicted under the assumption of local thermodynamic equilibrium given assumptions of the column density, excitation temperature, systemic velocity, line width and source size. Typically besides the main (very optically thick) lines the three latter parameters can be fixed for all lines and species leaving the column densities to be constrained. 

To decide which molecules to assign we compared our L483 data directly to those from the ALMA Protostellar Interferometric Line Survey (PILS) of the low-mass Class 0 protostellar binary IRAS~16293-2422 \mbox{\citep{2016A&A...595A.117J}}. We used the spectrum from the B component of the protostellar binary (IRAS16293B in the following). Specifically, in that survey more than 10,000 separate features can be identified  toward IRAS16293B in a frequency range between 329~GHz and 363~GHz. With the large frequency coverage in the PILS data many species can be well-identified and the column densities and excitation temperatures constrained.  Given our smaller frequency coverage in the observations of L483 the assignments would in their own right only be tentative and the inferred column densities mainly a sanity check that the assignments are plausible. The general agreement with the identifications in the IRAS16293B data strengthens this case, however. Table~\ref{transitions} lists the assigned transitions, while Table~\ref{columns} gives the inferred column densities for L483 and IRAS16293B.

For most species, we do not have a sufficient number of transitions to constrain the excitation temperature, except for CH$_2$DOH (deuterated methanol) and CH$_3$OCHO (methyl formate). For the former, an excitation temperature of $100\pm 25$~K reproduces the relative line strengths, while  the lines of CH$_3$OCHO are slightly better fitted with a higher excitation temperature of 300~K. This  situation is similar to that of IRAS16293B where a number of species, including methyl formate, with binding energies of 5000--7000~K are  best fitted with a high excitation temperature of a few hundred K, while other species require a lower excitation temperature of about 100~K.  Toward IRAS16293B, optically thin transitions of methanol is also best fitted with an excitation temperature of 300~K, but toward that source a colder component is also present as witnessed by extended emission in a number of lower excited transitions as well as the low temperatures of highly optically thick transitions. Toward L483, a number of the stronger lines of CH$_2$DOH with low upper energy levels ($\sim 100$~K) are marginally optically thick with $\tau$ of 0.1--0.5. Thus, it is plausible that a still higher temperature component with a high column density may be present on even smaller scales, not traced by the lines identified here. Contrary to the methanol isotopologue transitions, most of the methyl formate lines have low opacities of 0.01--0.05, and thus are very likely sampling the most compact, high column density material. Also, it should be noted that the inferred column densities are only weakly dependent on the exact excitation temperature, changing by less than $10$--$20$\% with temperatures varying from about $100$~K to 300~K. For the purpose of this paper, these uncertainties are less critical.

\begin{figure*}
 \includegraphics[scale=1.4]{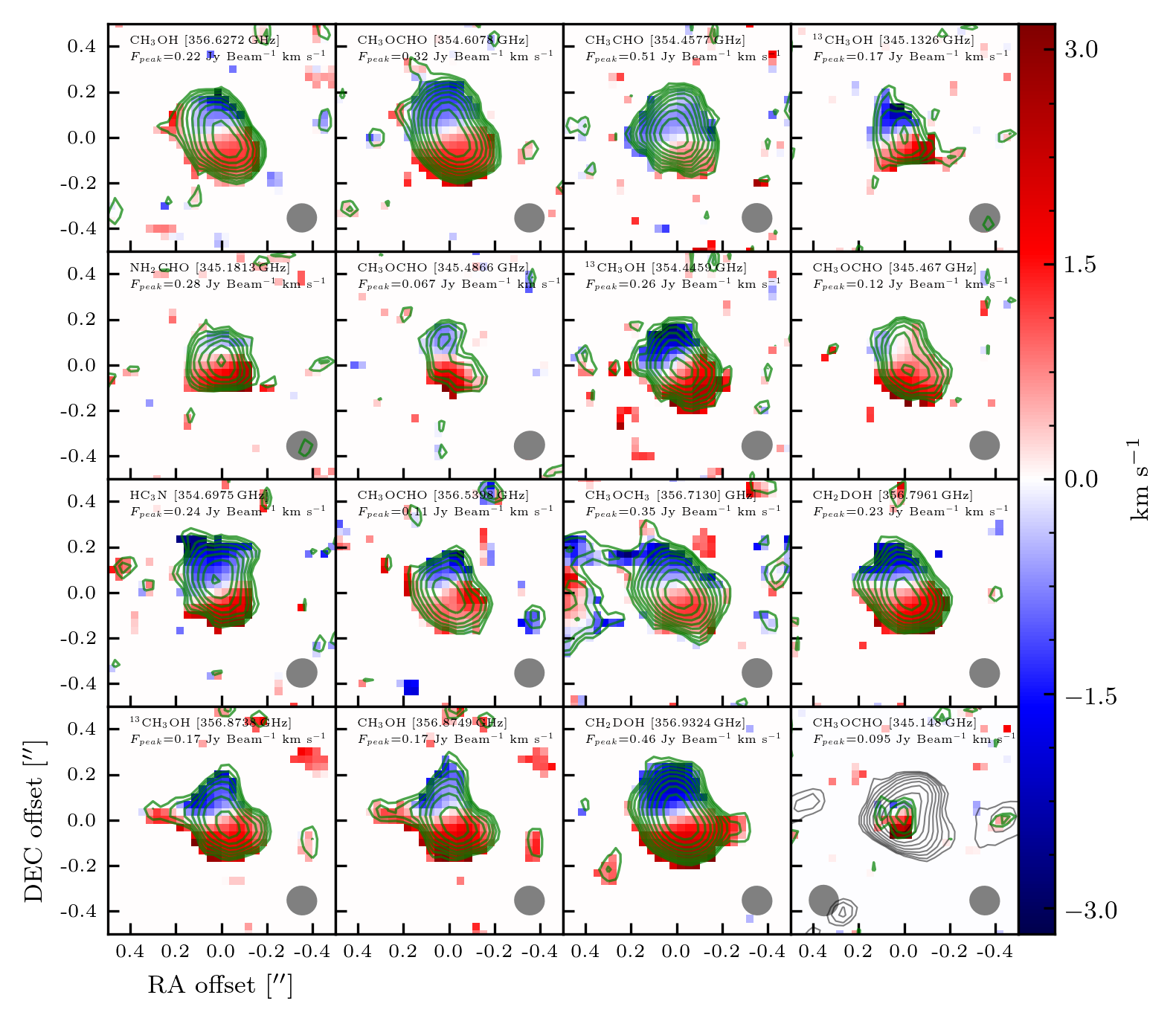}
 \caption{Moment~0 and 1 maps of observed COMs (Table~\ref{transitions}). Moment~0 maps are in green contours, overlaid on the moment~1 map. The Cycle~3 857~$\upmu$m dust continuum image is shown in the lower right frame, in gray contours, and the beamsize of the dust continuum observation is shown in the lower left corner of the same frame. Both dust continuum and the moment~0 map contours are spaced logarithmically between $5$--$100$~\% of the peak emission, in 10 steps. The first two panels show the unblended lines of CH$_3$OH and CH$_3$OCHO, while the remainder are blended to different degrees. The beamsizes of the molecule observations are shown in the lower right corner of each frame. The peak flux of the observed transition, $F_{\mathrm{peak}}$, is given in each panel, while the mean rms is 0.016~Jy~Beam$^{-1}$~km~s$^{-1}$. All the data shown in this figure are from Cycle~3.}
 \label{fig:sel_lines}
\end{figure*}

\FloatBarrier
We constructed moment~0 and 1 maps of each individual identified COM transition  in Fig.~\ref{fig:sel_lines}, integrating all emission within $\pm$~7~km~s$^{-1}$ based on Fig.~\ref{fig:imfit}. A range of COM molecules demonstrate line emission with clear velocity gradients similar to those seen in the four main species. Deconvolved 2D Gaussian fits to the moment~0 maps of the selected COMs reveal that most molecules are extended compared to the continuum peak with deconvolved sizes of approximately 0.2--0.3$''$ (40--60~au). 

\subsection{Comments about individual species/transitions}
Methanol (CH$_3$OH) is the most prominent organic molecule identified
on small scales towards solar-type protostars with different rarer
isotopologues typically possible for identification. In our frequency range,
the main lines of CH$_3$OH are of the two isotopologues,
$^{13}$CH$_3$OH and CH$_2$DOH. CH$_2$DOH, in particular, has six
transitions in the HCO$^+$ spectral window and cover two others in the
H$^{13}$CN and CS windows with three separate
transitions of $^{13}$CH$_3$OH. In the HCO$^+$ window, a prominent feature is seen at 356.625~GHz. 
The best option for this line is a set of relatively highly excited
transitions of the main isotopologue of methanol (CH$_3$OH
$23_{-4}-22_{-5}$). For the derived column density of $^{13}$CH$_3$OH, these
transitions should indeed be present at a temperature of $100$~K taking into account
the standard $^{12}$C:$^{13}$C ratio of 68 \citep{2005ApJ...634.1126M}. A similar highly excited
CH$_3$OH transition at 356.875~GHz is blended with a
transition of $^{13}$CH$_3$OH.  Methyl mercaptan (CH$_3$SH) at 356.627~GHz ($E_u = 136$~K; $\log_{10} A_{ul} = -4.0$ [s$^{-1}$]) could also contribute to this feature. However, unless its
excitation would be very peculiar, one would then also expect to
see CH$_3$SH transitions at 354.372~GHz  ($E_u = 147$~K; $\log_{10} A_{ul} = 3.4$ [s$^{-1}$]) and
354.643~GHz  ($E_u = 146$~K; $\log_{10} A_{ul} = -3.4$ [s$^{-1}$]) in the HCN window with approximately the same
strengths. CH$_3$SH is indeed detected toward IRAS16293B \citep{drozdovskaya18} in the PILS data with the three transitions at 356.625, 354.372, and 354.643~GHz all clearly seen. As the latter two do not show up toward L483 it appears that CH$_3$SH does not contribute at this level. 

A few isolated transitions can be assigned to individual species with an assumed excitation temperature of $100$~K. These include CH$_3$OCH$_3$, C$_2$H$_5$OH, NH$_2$CHO, SO$_2$, H$_2$CS, and HC$_3$N. Of these species, the relatively common gas phase molecules, H$_2$CS, SO$_2$, and HC$_3$N, are found to be relatively more abundant toward L483 than the complex organics. For the other species, the inferred column densities are in agreement with those toward IRAS16293B, lending credibility to their assignments.

A few features remain problematic to assign. For example, one feature at 356.52~GHz
in the HCO$^+$ window could be attributed to a few different species, including
ethylene glycol and acetone, but these species would have transitions
visible in other spectral windows. In the PILS data, a feature is also seen
at this frequency, which is also not easily assigned to any of the
tabulated species. 

In the HCN window, the feature at 354.458~GHz is somewhat puzzling. By itself, it could be attributed to acetaldehyde (CH$_3$CHO) but this species has a similar transition at 354.525~GHz that should be equally strong and in the PILS data the two transitions in fact show up in this way. This behavior is noteworthy as acetaldehyde otherwise is considered one of the most easily identifiable of the complex organics, but in our data it can thus only be tentatively identified.
 
In the H$^{13}$CN window, the feature around 345.285~GHz remains
unassigned. It could be attributed to cyanamide (NH$_2$CN)  at 345.2869~GHz ($E_u = 138$~K; $\log_{10} A_{ul} = -4.1$ [s$^{-1}$]), which would be an
excellent fit and was recently identified in the PILS data by
\mbox{\cite{2018MNRAS.475.2016C}}. To reproduce the observed line strength, however,
a cyanamide to formamide (NH$_2$CN/NH$_2$CHO) ratio of
20 would be required, whereas all other interferometric measurements, as well as models,
have formamide being more abundant than cyanamide by an order of
magnitude or more. More likely, there remains an issue with spectroscopic
predictions for line intensities. For example, two transitions are seen in PILS
data at the same frequencies that are unassigned: one is likely
CH$_2$DOH (345.2842~GHz) and the other C$_2$H$_5$OH (345.2877~GHz). However, the B- and C-type transitions of deuterated methanol are known to be problematic and
previously some issues have been identified for C$_2$H$_5$OH as well
\mbox{\citep{2016A&A...587A..92M}}. For the feature at 356.546~GHz, the best assignment
would be of ethyl cyanide (CH$_3$CH$_2$CN). This species does have a
relatively bright transition in the HCN window at 354.477~GHz, but due to blending with 
the HCN transition itself it is not possible to see whether this transition is indeed present. Another
CH$_3$CH$_2$CN transition is near the HCO$^+$ window at 356.960~GHz, but falls just outside of
our frequency coverage. If the transition indeed could be solely attributed to CH$_3$CH$_2$CN, it would be more abundant by an order of magnitude relative to C$_2$H$_5$OH than what is seen toward IRAS16293B \citep{calcutt18}. Clearly, more transitions are needed to be observed of these species for reliable assignments and column densities.
\begin{table*}\centering
  \caption{Inferred column densities and fractional abundances relative to CH$_3$OH for the species identified toward
    L483 and compared to the one beam offset position from IRAS16293B \citep{2016A&A...595A.117J}.}\label{columns}
\label{tab:col_dens}
\begin{tabular}{llllll}\hline\hline
  \multicolumn{2}{l}{Species} & \multicolumn{2}{c}{Column density [cm$^{-2}$]} & \multicolumn{2}{c}{[X/CH$_3$OH]} \\
 & & L483 & IRAS16293B & L483 & IRAS16293B \\ \hline
Methanol         & CH$_3$OH         & $1.7\times 10^{19}$  & 1.0$\times10^{19}$ & 1 & 1 \\
                 & $^{13}$CH$_3$OH  & $2.5\times 10^{17}$  & 1.5$\times10^{17}$ & 1.5$\times10^{-2}$ & 1.5$\times10^{-2}$ \\ 
                 & CH$_2$DOH        & $4.0\times 10^{17}$  & 7.1$\times10^{17}$ & 2.4$\times10^{-2}$ & 7.1$\times10^{-2}$ \\
Dimethyl ether   & CH$_3$OCH$_3$    & $8.0\times 10^{16}$  & 2.4$\times10^{17}$ & 4.7$\times10^{-3}$ & 2.4$\times10^{-2}$ \\
Methyl formate   & CH$_3$OCHO       & $1.3\times 10^{17}$  & 2.6$\times10^{17}$ & 7.6$\times10^{-3}$ & 2.6$\times10^{-2}$ \\
Ethanol          & C$_2$H$_5$OH     & $1.0\times 10^{17}$  & 2.3$\times10^{17}$ & 5.9$\times10^{-3}$ & 2.3$\times10^{-2}$ \\
Acetaldehyde     & CH$_3$CHO        & $8.0\times 10^{16}$  & 1.2$\times10^{17}$ & 4.7$\times10^{-3}$ & 1.2$\times10^{-2}$ \\
Formamide        & NH$_2$CHO        & $1.0\times 10^{16}$  & 1.2$\times10^{16}$ & 5.9$\times10^{-4}$ & 1.2$\times10^{-3}$ \\
Cyanopolyyne     & HC$_3$N          & $5.0\times 10^{17}$  & 1.8$\times10^{14}$ & 2.9$\times10^{-2}$ & 1.8$\times10^{-5}$ \\
Thioformaldehyde & H$_2$CS          & $2.0\times 10^{16}$  & 1.5$\times10^{15}$ & 1.2$\times10^{-3}$ & 1.5$\times10^{-4}$ \\
Sulfur-dioxide   & SO$_2$           & $1.0\times 10^{17}$  & 1.5$\times10^{15}$ & 5.9$\times10^{-3}$ & 1.5$\times10^{-4}$ \\
\hline
\end{tabular}
\end{table*}
 
\section{Analysis}
\label{sec:anal}
\subsection{Kinematics}
\label{sec:kin}
To determine whether or not a Keplerian disk is present in L483, we investigated the kinematics of the gas motions in the inner 30~au of L483. For this purpose, we fit the position of the peak emission in each spectral cube channel, using the 2D Gaussian fit routine, CASA \texttt{imfit}. H$^{13}$CN~$J$~=~4--3 and CS~$J$~=~7--6, which both show signs of a rotation profile perpendicular to the outflow direction, were fit, while we do not consider HCN~$J$~=~4--3 and HCO$^{+}$~$J$~=~4--3 as these transitions are heavily influenced by the East-West outflow (Fig.~\ref{fig:mom_maps}). We do not attempt to constrain the velocity structure in the outflow directions, in this work. While CS~$J$~=~7--6 does contain extended emission, the \texttt{imfit} results near the continuum center in Fig.~\ref{fig:mom_cont_cs} show a clear velocity profile not visibly affected by the outflows in the image maps. For this reason we include CS~$J$~=~7--6 in our kinematic analysis. We further defined a box around the central emission as input to \texttt{imfit} to circumvent the large-scale emission seen in the South-East direction (Fig.~\ref{fig:mom_maps}) of H$^{13}$CN~$J$~=~4--3 and CS~$J$~=~7--6. This box region can be seen in Fig.~\ref{fig:mom_cont_h13cn}.

\begin{figure*}
 \includegraphics[width=\textwidth]{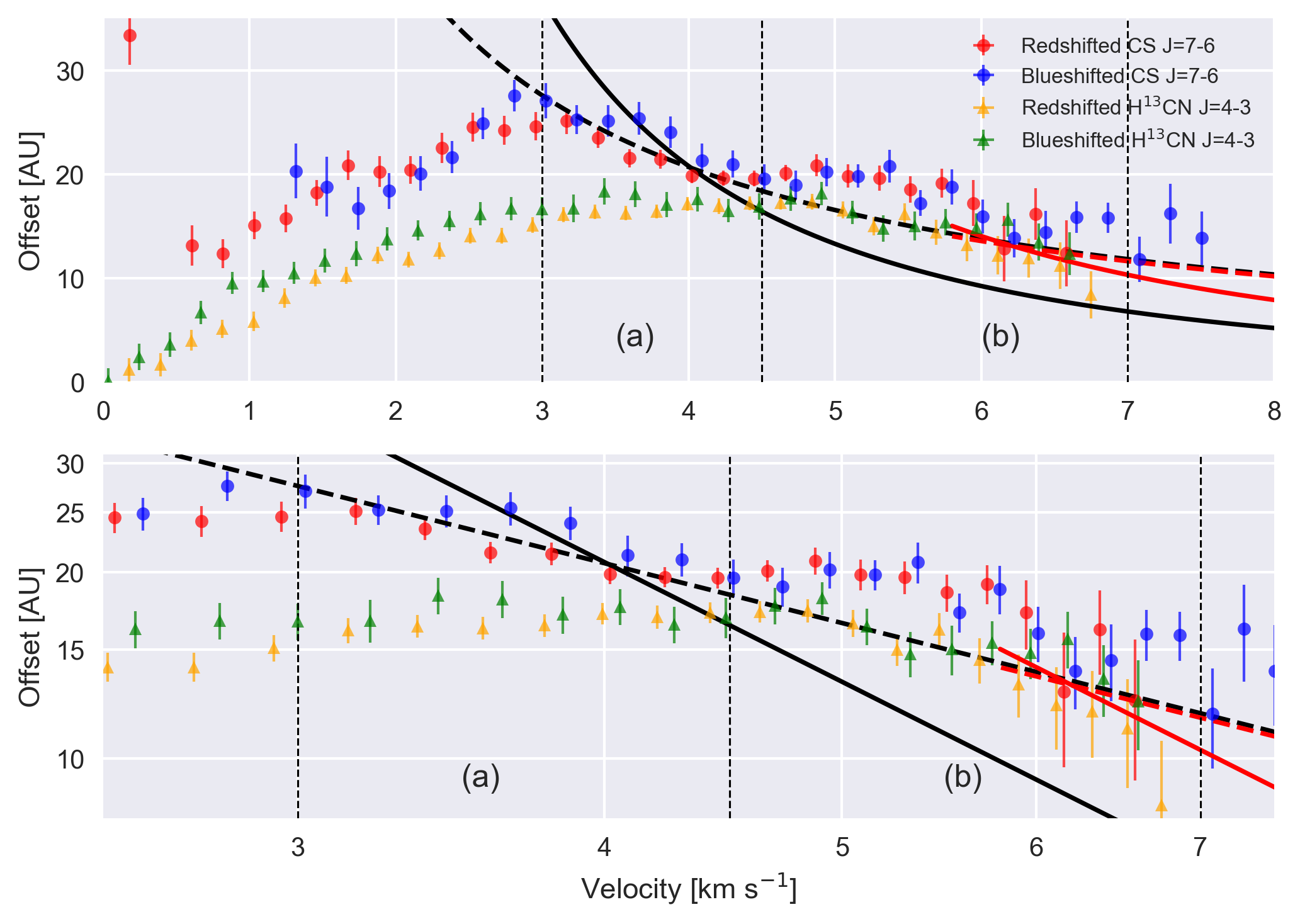}
 \caption{Distance vs. velocity plot of H$^{13}$CN~$J$~=~4--3 and CS~$J$~=~7--6 \texttt{imfit} datapoints, using Cycle~3 data only, in linear scale (top frame) and logscale (bottom frame). The velocities on the first axis are the velocity offset from the employed systemic velocity (6.0 km s$^{-1}$). The blue- and redshifted \texttt{imfit} datapoints of the \CS emission are shown in blue and red, respectively, while the blue- and redshifted \texttt{imfit} datapoints of \HthirtCN are shown in green and orange, respectively. The vertical bars of each datapoint represent the 1$\sigma$ uncertainty. The (a) and (b) regions, with borders marked by vertical dashed lines, show which datapoints were used for the fit. Only \CS datapoints are used for the fit in velocity region (a), since \HthirtCN data at velocities below region (b) appear affected by the projected velocities from the emission edge. Only \HthirtCN datapoints are used in velocity region (b), as the \CS emission appear affected by outflows in this region. The data diverging from the model velocity profiles at low velocities, below velocity region (a), arise from the projected rotational gas motions at the edge of the emission region, and are not used in the fit. Some \CS \texttt{imfit} results are lacking for v~$<1.25$~km~s$^{-1}$ due to heavy, global absorption at low velocities. Both frames are overlaid with the best fit infall (black dashed line) and Keplerian velocity profile (black full line) to the Cycle~3 data. The red solid line and the red dashed line show the best fit to the \HthirtCN \texttt{imfit} datapoints with absolute velocities >~$5.8$~km~s$^{-1}$, using a Keplerian and infall velocity profile, respectively.} 
 \label{fig:imfit}
\end{figure*}

For the two transitions, the \texttt{imfit} data points are mostly found in two ``clusters'', with  each cluster corresponding to blue- and red-shifted emission, respectively (see Fig.~\ref{fig:mom_cont_h13cn} and \ref{fig:mom_cont_cs}). We made a weighted linear regression fit to these clusters of data, which defines the velocity vector, and took the weighted average declination coordinate of each of the two data clusters as input to the inverse velocity vector function to define the average right ascension coordinate. The midpoint between these two representative points of the data clusters was taken as the emission center, determined to be $\alpha_{2000}$=18$^{\mathrm{h}}$17$^{\mathrm{m}}$29.943$\pm$0.001$^{\mathrm{s}}$, $\delta_{2000}$=-04$^{\circ}$39$^{\prime}$39$^{\prime\prime}$.595~$\pm$~0.012$^{\prime\prime}$.

The rotation axis was then defined as being normal to the velocity vector, intersecting the emission center, see Fig.~\ref{fig:mom_cont_h13cn}. We defined the velocity vector, rotation axis, and emission center using H$^{13}$CN~$J$~=~4--3 data and applied it to CS~$J$~=~7--6, as H$^{13}$CN~$J$~=~4--3 has more datapoints. The rest velocity for L483 was estimated to be 6.0~km~s$^{-1}$, using the H$^{13}$CN~$J$~=~4--3 and CS~$J$~=~7--6 emission, under the criterion that the red- and blueshifted datapoints should be symmetric in a distance vs. velocity plot, see Fig.~\ref{fig:imfit}. This is done under the assumption of optically thin emission. This rest velocity is used as the reference for the velocity offset of the \texttt{imfit} data points, while the data point distance from the rotation axis was taken as the offset distance. 

The \texttt{imfit} data points were then converted into (radius, velocity) points and used in a reduced $\chi^2$ fit of two different velocity profiles; a Keplerian velocity profile, $v = \sqrt{GM_*/r}$, where $G$ is the gravitational constant and $M_*$ is the central mass, and the velocity profile of infall with conservation of angular momentum, $v_\phi r=v_{\phi, 0}r_0$, where $v_{\phi, 0}$ and $r_0$ is the start velocity and starting distance, respectively, of the collapsing material. $r_0$ can be interpreted as the starting position of the given material in the cloud, at the time when the collapse started. Such an infall scenario will have $v_\phi \propto r^{-1}$, a velocity profile previously used to estimate the kinematics of a disk or disk-like structure around Class~0/I objects \mbox{\citep[e.g., see][]{2014A&A...566A..74L}}. For the Keplerian profile, we used a central mass range of 0.01-1.5~M$_{\odot}$ and $2.8\times10^5$ uniform steps in the specific angular momentum range for the infall velocity profile of 10$^4$ -- 1.5$\times10^5$ m$^2$ s$^{-1}$, to find the best fit. The central mass parameter presents M$_*$sin$i$, where $i$ is the system inclination, since we only observe radial velocities. Using an inclination between 75--90$^{\circ}$ \citep{2018ApJ...863...72O}, we can in principle extract an estimate of the true stellar mass from the best fit central stellar mass. However, the paramount goal is to distinguish between the two velocity profiles, not to obtain a precise measure of the stellar mass.

We combined the H$^{13}$CN~$J$~=~4--3 and CS~$J$~=~7--6 data into a single dataset, and performed a reduced $\chi^2$ fit in the velocity regime. We excluded the lowest velocities as these are affected by the small extent of the H$^{13}$CN gas line emission ($\sim$~20~au~radius), causing the lower velocities to be dominated by low-velocity gas toward the emission center, arising from projected gas velocities coming from the edge of the observed gas emission region. These low radial velocities appear at low projected distances, while the true distance to the protostar is unknown. As such, the emission seen at low projected distances in Fig.~\ref{fig:imfit} for $v<4.5$~km~s$^{-1}$, likely arise from larger actual distances to the protostar. Moreover, emission at these low velocities could include emission from the other disk-half, as thermally broadened lines from the other disk-half are convolved with the relatively large beam, drawing the \texttt{imfit} results towards a lower offset distance. We also excluded higher velocities, where the CS~$J$~=~7-6 \texttt{imfit} results become noisy. The cause of this noise is unknown, but it could be an effect of high-velocity outflowing material. The difference in the spatial extents of the observed CS and H$^{13}$CN datapoints and the spatial disparity illustrated in Fig.~\ref{fig:mom_maps}, are likely related to the lower critical density of CS relative to HCN \mbox{\citep{1999ARA&A..37..311E}}.

This approach of using the emission peak position in each channel to constrain the gas kinematics, is only exactly valid when each emission component comes from a single position. If this is a poor approximation of the true velocity structure in each velocity component, then systematic uncertainties will be introduced. However, given the systematic and concentrated peak emission positions of the gas line emission velocity components of \CS and \HthirtCN in Fig.~\ref{fig:mom_cont_h13cn} and \ref{fig:mom_cont_cs}, this approach should be meaningful without severe systematic uncertainties. 

We used both the combined dataset and Cycle~3 data alone in the reduced $\chi^2$ fit, as the \texttt{imfit} result precision is affected negatively by the larger rms and beam in the combined dataset. The Cycle~3 data have more precise \texttt{imfit} data points, due to their higher angular resolution of $\sim0.13^{\prime\prime}\times0.13^{\prime\prime}$ and lower rms, while missing the shorter baselines of the combined Cycle~1 and 3 data. 
Consequently, spatial filtering is seen in the Cycle~3-only continuum emission in Fig. \ref{fig:mom_cont_h13cn} and Fig.~\ref{fig:mom_cont_cs} when compared to the Cycle~1+3 continuum emission in Fig. \ref{fig:dust_cont}, even after taking into account the different beam-sizes.
Fig.~\ref{fig:imfit} shows that, using Cycle~3--only data, the infall profile is strongly favored by the reduced $\chi^2$ fit, with $\chi^2_{\mathrm{red}}$~=~0.84, while the Keplerian fit has $\chi^2_{\mathrm{red}}=11.6$. Using the combined dataset gave the same conclusion: the infall profile is heavily favored, with $\chi^2_{\mathrm{red}}$ better by more than a factor of five. This result suggests that the observed line emission is not from gas in a rotationally supported disk, but rather from gas in an infalling-rotating structure.
While the CS~$J$~=~7--6 emission enabled the kinematic analysis to be extended to larger radii, concerns exists whether or not it could be influenced by the outflows, as extended CS~$J$~=~7--6 is seen in the moment maps in Fig.~\ref{fig:mom_maps}. For this reason, a reduced $\chi^2$ fit was also made on the combined and Cycle~3 data using only H$^{13}$CN~$J$~=~4--3 emission data. The conclusion remain the same, with infall favored for both the combined dataset and Cycle~3-only data. 
In the event that a Keplerian disk exists, we would expect a transition between an infalling velocity profile and a Keplerian velocity profile, at the disk edge. We do see a tentative change in the redshifted \HthirtCN \texttt{imfit} data near $5.8$~km~s$^{-1}$, at $\sim$~$15$~au, so we performed an independent fit to \HthirtCN Cycle~3 data above this velocity, seen in Fig.~\ref{fig:imfit}. The reduced $\chi^2$ fit slightly favored a Keplerian velocity profile with $\chi^2_{\mathrm{red}}$=$0.71$ vs. an infall velocity profile with $\chi^2_{\mathrm{red}}$=$0.85$. However, we cannot conclude the presence of a Keplerian disk with a radius of $15$~au, since the data is too noisy and sparse, and since an infall velocity profile could fit the data as well.
The absence of a Keplerian disk down to at least $15$~au is consistent with the analysis of the submillimeter continuum emission toward the source \mbox{\citep{2004A&A...424..589J,2007ApJ...659..479J, 2009A&A...507..861J}} where the interferometric flux of L483 was consistent with envelope-only emission and did not need a central compact emission source. 

\subsection{Dust temperature profile and the distribution of COMs}\label{sec:temp}
\label{sec:kin_coms}
In order to investigate the dust temperature profile of L483 on small scales, we used the density solution to an infalling-rotating collapse \citep{1984ApJ...286..529T}, with an example centrifugal radius of $60$~au \citep[consistent with the conclusions of][]{2017ApJ...837..174O}. For an initial total dust mass guess, we performed a 2D Gaussian fit on the $857$~$\upmu$m emission (Fig.~\ref{fig:dust_cont}), extracted from a box around the elongated emission structure and found a deconvolved size of $0.36\arcsec\times0.20\arcsec$. 
The box was chosen to avoid the elongated emission in the East-West direction, and instead focus on the inner 60--80 au, where the bulk emission is present.
We approximate the deconvolved fit to a circular region, with an extent of 0.28\arcsec, i.e., we approximate the observed dust continuum to a spherical model of 56~au radius. While the $857$~$\upmu$m dust continuum traces material swept up in the outflow structure, we do not attempt to model the outflow or outflow cavities. Using the available SED data (Table~\ref{tab:sed}), we integrated the SED and estimate the bolometric luminosity to be 10.5~L$_{\odot}$, comparable with previous luminosity estimations of $9$~L$_{\odot}$ \mbox{\citep{2002A&A...389..908J}} and $13$~$\pm$~$2$~L${_\odot}$ \citep{2000ApJS..131..249S}.

The total dust mass is given by 
\begin{equation}
M = \frac{S_\nu d^2}{\kappa_\nu B_\nu(T)},
\label{eq:mass}
\end{equation}
where $S_\nu$ is the total source flux, $d$ is the distance, $\kappa_\nu$ is the dust opacity, and $B_\nu(T)$ is the spectral radiance.
Both $\kappa_\nu$ and $B_\nu(T)$ depend on the temperature field as the mean dust opacity $\kappa_\nu$ will be a mixture of dust with and without ice-mantles due to sublimation caused by heating from the central protostar. We used bare-grain and thin ice-mantle opacities from \mbox{\citet{1994A&A...291..943O}}, corresponding to coagulated dust grains in an environment with a gas number density of $10$$^6$~cm$^{-3}$. We used initial guesses of $T_\mathrm{av}$~=~100~K, the mean dust temperature of all dust, both with and without ice mantles, within 56~au, and a dust population ratio between icy-dust and bare-grain dust of 0.5, also within 56~au, to get an initial estimate of the total dust mass, in the innermost region of L483, using Eq.~\ref{eq:mass}.

\begin{figure*}
 \includegraphics[width=\textwidth]{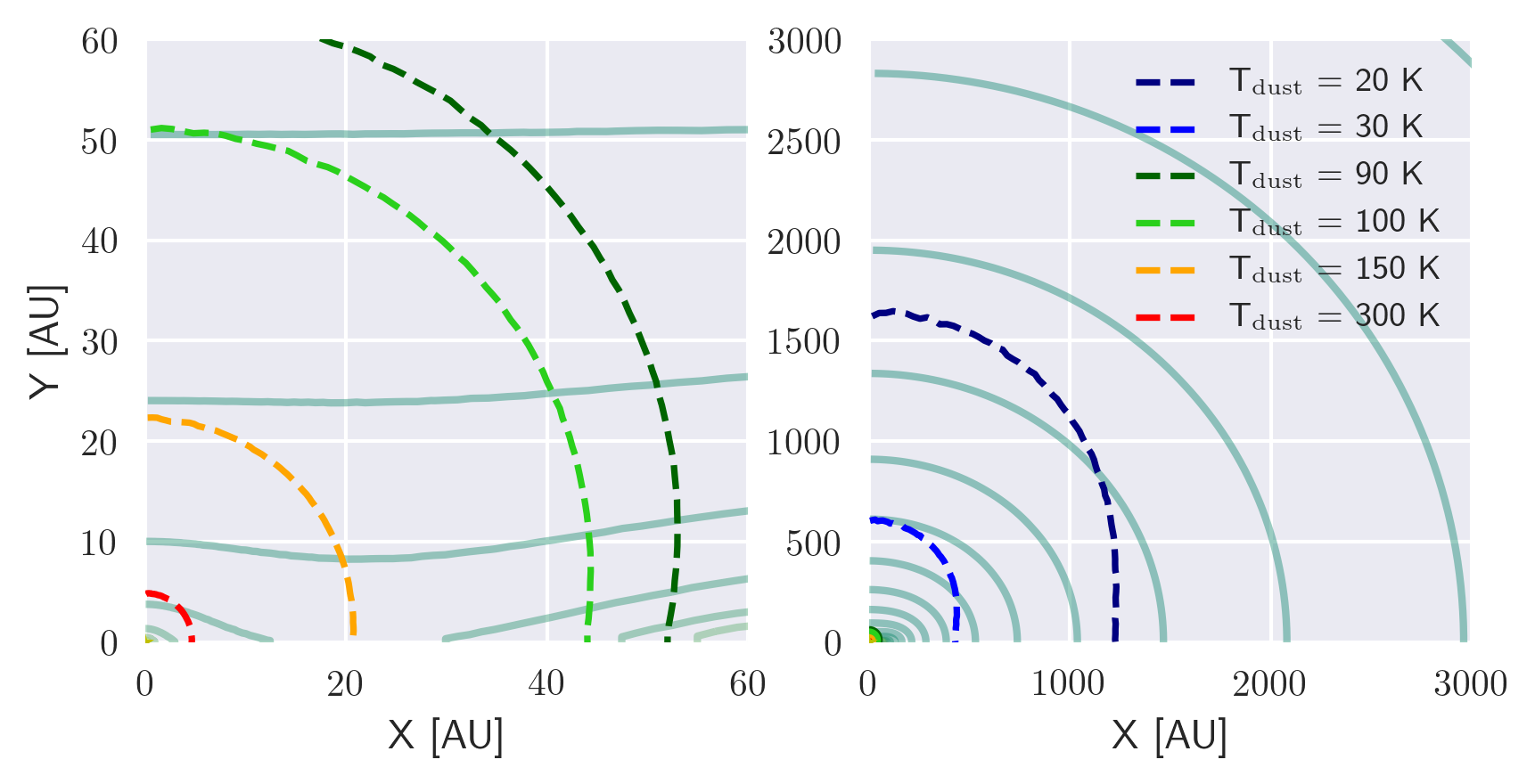}
 \caption{Temperature distribution in the density solution to an infalling-rotating collapse model, with $r_c$~=~$60$~au, $M_{*}$~=~$0.2$~M$_{\odot}$, and $\dot{M}$~=~$4.05\times10^{-7}$ M$_\odot$~yr$^{-1}$ using only coagulated bare-grain opacities, assuming a gas number density of 10$^6$ cm$^{-3}$. The density contours in solid lines are logarithmically spaced, increasing towards the midplane (the x-axis), while the dust temperature contours are in dashed lines.} 
 \label{fig:temp_dust_bare}
\end{figure*}

With a dust mass estimate as input, we used \texttt{RADMC-3D}, a 3D Monte Carlo radiative transfer code \citep{2012ascl.soft02015D}, to determine the dust temperature, which led to a new mass estimation, as $T_\mathrm{av}$ and $\kappa_\nu$ change (Eq.~\ref{eq:mass}). The updated mass in turn led to a different temperature distribution, which again affects our estimate of $\kappa_{\nu}$ and $T_{\mathrm{av}}$. After a few iterations, we had a stable estimate of all parameters, with $T_{\mathrm{av}}$=$125$~K from visual inspection of the temperature distribution within 56~au (Fig.~\ref{fig:temp_dust_bare}). Almost all the dust within 56~au has temperatures above $90$--$100$~K,  leading us to adopt bare-grain dust opacities exclusively, as the water-ice mantle sublimates at these temperatures \citep{1993ApJ...417..815S}.
The final temperature profile can be seen in Fig.~\ref{fig:temp_dust_bare} and we estimate the total mass (dust + gas) in the inner region to be $8.8\times10^{-4}$~M$_{\odot}$, using a gas-to-dust mass ratio of 100. While the exact temperature distribution and derived total mass using Eq.~\ref{eq:mass} is dependent on the used dust density model and its parameters, we have used a dust density model consistent with both our observed kinematics (the gas kinematics of the model has a $v_\phi \propto r^{-1}$ profile) and the earlier research of \citet{2017ApJ...837..174O}. We also re-performed the continuum analysis and radiative transfer modeling using opacities of bare-grains coagulated with higher gas number densities of 10$^7$ and 10$^8$ cm$^{-3}$, which showed a consistent, but minor drop in sublimation radius as the ambient gas density of the coagulated bare-grain opacities increased. The 44-52 au sublimation radius of the $10^6$ cm$^{-3}$ gas density model is decreased to a 41-49 au sublimation radius in the $10^8$ cm$^{-3}$ gas density model. Since changing the model opacity also reduces the required amount of material in the model (Eq.~\ref{eq:mass}), the total optical depth of the model is not significantly changed, and the sublimation radius is therefore not dramatically affected by different opacity models.

The spatial extents of the COMs (Fig.~\ref{fig:sel_lines}) were fitted with 2D Gaussian profiles, from which the deconvolved major and minor axes were found. The mean major axis is 0.2$\arcsec$ with a $2\sigma$ deviation of 0.1$\arcsec$. The maximal spatial extents of the COMs emission (0.2--0.3$^{\prime\prime}$, i.e., 40--60 au) are consistent with the estimated ice-mantle sublimation front of $\sim$~$50$~au (Fig.~\ref{fig:temp_dust_bare}), implying that the COMs reside in the hot corino, which is dominated by rotational motion. 

\begin{figure*}
 \includegraphics[width=\textwidth]{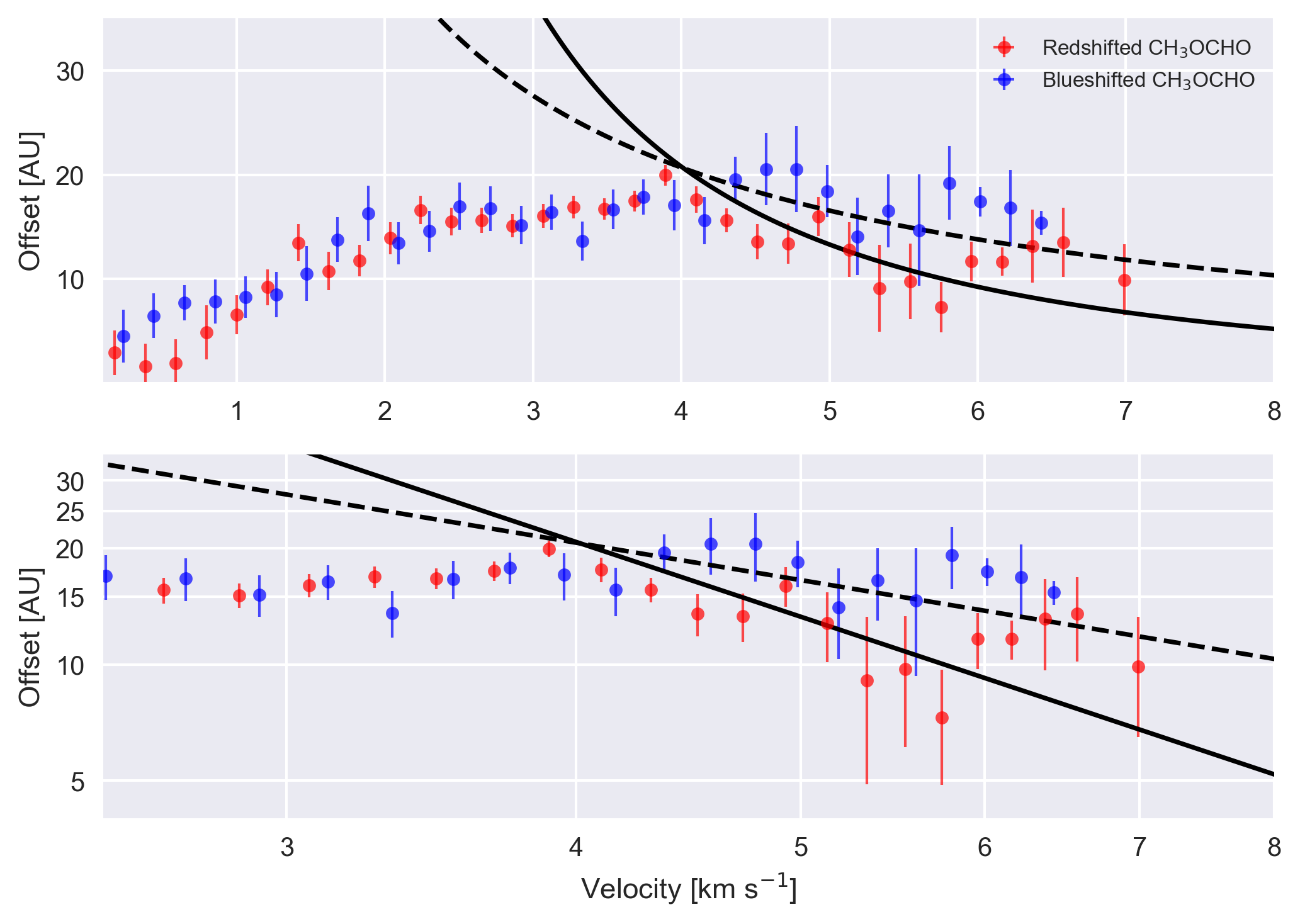}
 \caption{Distance vs. velocity plot of CH$_3$OCHO \texttt{imfit} datapoints, using Cycle~3 data only, in linear scale (top frame) and logscale (bottom frame). The velocities on the first axis are the velocity offset from the employed systemic velocity (6.0 km s$^{-1}$). The blue- and redshifted \texttt{imfit} datapoints of the CH$_3$OCHO emission are shown in blue and red, respectively. The data diverging from the model velocity profiles at low velocities, below $\sim$~$4$~km~s$^{-1}$, arise from the projected rotational gas motions at the edge of the emission region. Both frames are overlaid with the best fit infall (dashed line) and Keplerian velocity profile (full line) to the Cycle~3 \CS and \HthirtCN data.} 
 \label{fig:imfit_COM}
\end{figure*}

Figure~\ref{fig:sel_lines} shows that rotation profiles are also observed for all strong COM lines, with the same North-South velocity gradient as seen for CS~$J$~=~7--6 and H$^{13}$CN~$J$~=~4--3 (Fig.~\ref{fig:mom_maps}). We extracted the peak emission position in each channel for all observed COM lines in the Cycle~3 data using \texttt{imfit}, but the data were too noisy and the lines too blended for clear velocity profiles to be extracted, except for CH$_3$OCHO. We defined the emission center and rotation axis of CH$_3$OCHO based on its \texttt{imfit} data, as the CH$_3$OCHO emission is slightly offset compared to H$^{13}$CN~$J$~=~4--3. This offset is, however, well within the beam. Fig.~\ref{fig:imfit_COM} shows that the velocity profile of CH$_3$OCHO is consistent with the infall profile derived from the CS~$J$~=~7--6 and H$^{13}$CN~$J$~=~4--3 Cycle~3 data. 

The derived column densities of the COMs (Table~\ref{tab:col_dens}) are also in good agreement with the values found for COMs in IRAS16293B both in an absolute and relative sense. That there is a good agreement in an absolute sense, is likely because the amount of material on small scales at high temperatures is comparable toward the two sources. 
The envelope masses are also similar ($4.4$~M$_\odot$ vs. $4$~M$_\odot$) and while L483 is more luminous than IRAS16293B ($10.5$~L$_\odot$ vs. $\sim$3~L$_\odot$), IRAS16293B is surrounded by massive amounts of disk-like material \mbox{\citep{jacobsen18a}}, which may explain the comparable amount of material at high temperatures. The sulfur-species and HC$_3$N show higher column densities toward L483 than IRAS16293B, but whether this is a chemical effect, or rather reflect different physical structures on these scales, is unclear.

\section{Discussion}
\label{sec:disc}
The lack of an observable rotationally supported disk in L483 down to $\sim$$15$~au, and the presence of the COMs on scales of $40$--$60$~au, have a number of important implications. In terms of the system geometry, the position angle of our rotation profile was 11$^{\circ}$, perpendicular to the  outflow in the East-West direction. Our data show that L483 is undergoing a rotating collapse, with a large-scale outflows consistent with earlier works and a small inner region dominated by rotational motion (Fig.~\ref{fig:sel_lines}).
The infalling-rotating collapse continues down to at least $15$~au in radius, with the outflows of L483 necessarily being launched very close to the central protostar, in the absence of a disk >~$15$~au.
 
L483 shows some similarities with the Class~0 object B335, as both YSOs have a lack of Keplerian disks, $>$$15$~au and $>$$10$~au, respectively, and show very small amounts of dust in the innermost region, 8.8$\times10^{-4}$~M$_{\odot}$ vs. 7.5$\times10^{-4}$~M$_{\odot}$, for L483 and B$335$ \citep{2015ApJ...814...22E}, respectively. Both objects contain a hot corino \citep{2016ApJ...830L..37I}, while the L483 hot corino radius of $40$--$60$~au is likely larger than that of B$335$, estimated to be only a few tens of au \citep{2016ApJ...830L..37I}, due to its relatively low luminosity of 0.72~L$_\odot$ \citep{2015ApJ...814...22E}. B$335$, however, has low levels of rotation in its infall, whereas L483 has a clear infall-rotational signature (Figs.~\ref{fig:mom_maps} and \ref{fig:sel_lines}).

\citet{2017ApJ...837..174O} invoke a chemical transition at the centrifugal barrier of L483, due to the abrupt transition in the physical environment. They also speculate that NH$_2$CHO and CH$_3$OCHO reside in an unresolved Keplerian disk, explaining the compact emission they observe.
Our data of emission from the same molecules and the absence of a Keplerian disk down to a radius of at least $15$~au, however, illustrate that these species also reside in the infalling envelope. Also, these COMs, together with the distributions of the other COMs, can be accounted for by the release of molecules into the gas phase due to dust ice-mantle sublimation by itself. They invoked a Keplerian disk model inside the centrifugal barrier to explain the compact, high-velocity emission structure they observe out to $\pm~$6~km~s$^{-1}$ in their PV diagrams, as their rotating-collapse model alone could not explain this emission together with the more spatially extended emission. However, they do not resolve the hot corino region ($0.2^{\prime\prime}$--$0.3^{\prime\prime}$) in their observations. We find empirically, using the peak emission position in each channel, that the compact high-velocity \CS and \HthirtCN emission up to at least $\pm$~$5.8$~km~s$^{-1}$, is best matched by an infalling velocity profile, not a Keplerian one.

L483, interestingly, also exhibits a difference to some of the more evolved protostars with larger disks. For example, \mbox{\citet{2014A&A...566A..74L}} find very small levels of CH$_3$OH toward the Class~$0$/I~protostar R~CrA-IRS$7$B and the presence of a Keplerian disk around this source. Based on a detailed line radiative transfer analysis, they demonstrate that this lack of CH$_3$OH emission may reflect the low column density of the protostellar envelope at the scales where material is being assembled into the circumstellar disk. A similar situation is seen toward the Class~I protostar Oph-IRS67 in Ophiuchus. For this source, \citet{artur18} find a Keplerian disk to be present and also do not see signs of any CH$_3$OH down to low column densities. In contrast, acetaldehyde was found towards the Class~$0$ object HH212 in Orion, where a tentative detection of a $\sim$~$90$~au Keplerian disk was made \citep{2014ApJ...786..114L}, though the abundance could not be determined due to optically thick submillimeter continuum emission \citep{2016A&A...586L...3C}. More observations are needed to quantify the relationship between the COM column density and abundance, and the presence of Keplerian disks.

For sources with an extended disk, the mass budget is dominated by the disk plane in the inner region. If the only source of heating at these scales is the radiation by the newly formed protostar, a significant amount of the material on small scales may be relatively cold, causing molecules to freeze-out onto dust grains and thus lowering the column density of COMs. In contrast, for YSOs with smaller disks, the warm envelope will dominate the mass budget on these scales.  As pointed out by \mbox{\citet{2014A&A...566A..74L}}, in the picture of a simple rotating collapse, an early stage hot corino without a sizable disk can only exist for a limited time: the radius of the region with the COMs being present in the gas-phase increases with the stellar mass as $R_{100~\mathrm{K}}$ $\propto M_*^{0.5}$, while the centrifugal radius within which rotational support is greater than the gas pressure \mbox{\citep{1984ApJ...286..529T}} grows as $r_c \propto M_*$. Thus, while the COMs can be abundant in the infalling envelope in the early stage where only a small Keplerian disk is present, the disk will grow more rapidly than the hot corino region and thus suppress its emission, as the COMs are trapped in ices in the disk midplane. The multitude of COMs in L483 and the lack of a Keplerian disk down to at least $15$~au, may therefore indicate that L483 is still in a chemical stage dominated by a warm inner envelope. In this picture, assuming that we have a protostar luminous enough to create an observable hot region, it is possible that these phenomena could be complementary, i.e., the presence of hot corino chemistry would signify the presence of a small disk. Similarly, sources with extended disks will show relatively small amounts of methanol and more complex species present in the gas phase. More observations relating the abundances of COMs to disk sizes, are needed to confirm this picture.

\section{Conclusion}
\label{sec:conc}
We have presented ALMA Cycles~1 and 3 Band~7 high angular resolution ($\sim$~$0.1^{\prime\prime}$) observations of HCN~$J$~=~4--3, H$^{13}$CN~$J$~=~4--3, CS~$J$~=~7--6, and HCO$^{+}$~$J$~=~4--3 together with of a series of complex organic molecules, towards the low-mass Class~0 object L483.
\begin{itemize}
  \item We fitted combined ALMA Cycles~$1$ and $3$ observations, as well as Cycle~$3$-only observations, of H$^{13}$CN~$J$~=~4--3 and CS~$J$~=~7--6 with two velocity profiles, Keplerian orbital motion and infall following angular momentum conservation. We found that the observed kinematics strongly favors an infall velocity profile. This result excludes the presence of a Keplerian disk in L483 down to a $\sim$~$15$~au radius. 
  \item A range of complex organic molecules was observed with the same rotational signature as H$^{13}$CN~$J$~=~4--3 and CS~$J$~=~7--6, from which a single clear velocity profile belonging to CH$_3$OCHO was extracted that follow the same infall profile as H$^{13}$CN~$J$~=~4--3 and CS~$J$~=~7--6.
  \item The emission of the observed complex organic molecules extends to $\sim$~40--60~au radius, consistent with the derived sublimation radius of $\sim$50~au, where the molecules sublimates into the gas phase off the dust grains, suggesting that the complex organic molecules exists in the hot corino of the L483 envelope. 
  \item The lack of a Keplerian disk down to at least a $15$~au radius, and the presence of complex organic molecules in the envelope at $\sim$~$40$--$60$~au scales, reveal that the complex organic molecules in L483 exists in a chemical era before the expected growing disk will dominate the chemistry and possibly reduce the column densities of complex organic molecules.
\end{itemize}

The ALMA telescope has facilitated observations of unprecedented angular resolution of low-mass protostars, revealing the innermost regions of the early star-forming stages. The exact timeline for the emergence of Keplerian disks, the chemistry of saturated complex organic molecules before and during the disk era remains poorly understood due to the low number of observations of such objects. With observations of more sources, the coming years will reveal the evolution of complex organic molecules and Keplerian disks on the smallest scales in nearby star-forming regions  and shed further light on their implications for the formation of planetary systems.

\begin{acknowledgements}
This paper makes use of the following ALMA data: 2012.1.00346.S and 2015.1.00377.S. ALMA is a partnership of ESO (representing its member states), NSF (USA) and NINS (Japan), together with NRC (Canada) and NSC and ASIAA (Taiwan) and KASI (Republic of Korea), in cooperation with the Republic of Chile. The Joint ALMA Observatory is operated by ESO, AUI/NRAO and NAOJ. 
S.K.J. and J.K.J. acknowledges support from the European Research Council (ERC) under the European Union's Horizon 2020 research and innovation programme (grant agreement No. 646908) through ERC Consolidator Grant ``S4F''. Research at the Centre for Star and Planet Formation is funded by the Danish National Research Foundation. This research has made use of NASA’s Astrophysics Data System.
This research made use of Astropy, a community-developed core Python package for Astronomy \cite{2013A&A...558A..33A}.
This work uses \texttt{PVEXTRACTOR}, see \url{https://github.com/radio-astro-tools/pvextractor}

\end{acknowledgements}
\bibliographystyle{aa} 
\bibliography{L483_coms,L483_coms_new} 

\begin{thebibliography}{62}
\expandafter\ifx\csname natexlab\endcsname\relax\def\natexlab#1{#1}\fi

\bibitem[{{Artur de la Villarmois} {et~al.}(2018){Artur de la Villarmois},
  {Kristensen}, {J{\o}rgensen}, {Bergin}, {Brinch}, {Frimann}, {Harsono},
  {Sakai}, \& {Yamamoto}}]{artur18}
{Artur de la Villarmois}, E., {Kristensen}, L.~E., {J{\o}rgensen}, J.~K.,
  {et~al.} 2018, A\&A in press. (arXiv:1802.09286)

\bibitem[{{Astropy Collaboration} {et~al.}(2013){Astropy Collaboration},
  {Robitaille}, {Tollerud}, {Greenfield}, {Droettboom}, {Bray}, {Aldcroft},
  {Davis}, {Ginsburg}, {Price-Whelan}, {Kerzendorf}, {Conley}, {Crighton},
  {Barbary}, {Muna}, {Ferguson}, {Grollier}, {Parikh}, {Nair}, {Unther},
  {Deil}, {Woillez}, {Conseil}, {Kramer}, {Turner}, {Singer}, {Fox}, {Weaver},
  {Zabalza}, {Edwards}, {Azalee Bostroem}, {Burke}, {Casey}, {Crawford},
  {Dencheva}, {Ely}, {Jenness}, {Labrie}, {Lim}, {Pierfederici}, {Pontzen},
  {Ptak}, {Refsdal}, {Servillat}, \& {Streicher}}]{2013A&A...558A..33A}
{Astropy Collaboration}, {Robitaille}, T.~P., {Tollerud}, E.~J., {et~al.} 2013,
  \aap, 558, A33

\bibitem[{{Bontemps} {et~al.}(1996){Bontemps}, {Andre}, {Terebey}, \&
  {Cabrit}}]{1996A&A...311..858B}
{Bontemps}, S., {Andre}, P., {Terebey}, S., \& {Cabrit}, S. 1996, \aap, 311,
  858

\bibitem[{{Bottinelli} {et~al.}(2004){Bottinelli}, {Ceccarelli}, {Neri},
  {Williams}, {Caux}, {Cazaux}, {Lefloch}, {Maret}, \&
  {Tielens}}]{2004ApJ...617L..69B}
{Bottinelli}, S., {Ceccarelli}, C., {Neri}, R., {et~al.} 2004, \apjl, 617, L69

\bibitem[{{Brinch} {et~al.}(2007){Brinch}, {Crapsi}, {J{\o}rgensen},
  {Hogerheijde}, \& {Hill}}]{2007A&A...475..915B}
{Brinch}, C., {Crapsi}, A., {J{\o}rgensen}, J.~K., {Hogerheijde}, M.~R., \&
  {Hill}, T. 2007, \aap, 475, 915

\bibitem[{{Calcutt} {et~al.}(2018){Calcutt}, {J{\o}rgensen}, {M{\"u}ller},
  {Kristensen}, {Coutens}, {Bourke}, {Garrod}, {Persson}, {van der Wiel}, {van
  Dishoeck}, \& {Wampfler}}]{calcutt18}
{Calcutt}, H., {J{\o}rgensen}, J.~K., {M{\"u}ller}, H.~S.~P., {et~al.} 2018,
  \aap, 616, A90

\bibitem[{{Chapman} {et~al.}(2013){Chapman}, {Davidson}, {Goldsmith}, {Houde},
  {Kwon}, {Li}, {Looney}, {Matthews}, {Matthews}, {Novak}, {Peng},
  {Vaillancourt}, \& {Volgenau}}]{2013ApJ...770..151C}
{Chapman}, N.~L., {Davidson}, J.~A., {Goldsmith}, P.~F., {et~al.} 2013, \apj,
  770, 151

\bibitem[{{Codella} {et~al.}(2014){Codella}, {Cabrit}, {Gueth}, {Podio},
  {Leurini}, {Bachiller}, {Gusdorf}, {Lefloch}, {Nisini}, {Tafalla}, \&
  {Yvart}}]{2014A&A...568L...5C}
{Codella}, C., {Cabrit}, S., {Gueth}, F., {et~al.} 2014, \aap, 568, L5

\bibitem[{{Codella} {et~al.}(2016){Codella}, {Ceccarelli}, {Cabrit}, {Gueth},
  {Podio}, {Bachiller}, {Fontani}, {Gusdorf}, {Lefloch}, {Leurini}, \&
  {Tafalla}}]{2016A&A...586L...3C}
{Codella}, C., {Ceccarelli}, C., {Cabrit}, S., {et~al.} 2016, \aap, 586, L3

\bibitem[{{Coutens} {et~al.}(2015){Coutens}, {Persson}, {J{\o}rgensen},
  {Wampfler}, \& {Lykke}}]{2015A&A...576A...5C}
{Coutens}, A., {Persson}, M.~V., {J{\o}rgensen}, J.~K., {Wampfler}, S.~F., \&
  {Lykke}, J.~M. 2015, \aap, 576, A5

\bibitem[{{Coutens} {et~al.}(2018){Coutens}, {Viti}, {Rawlings}, {Beltr{\'a}n},
  {Holdship}, {Jim{\'e}nez-Serra}, {Qu{\'e}nard}, \&
  {Rivilla}}]{2018MNRAS.475.2016C}
{Coutens}, A., {Viti}, S., {Rawlings}, J.~M.~C., {et~al.} 2018, \mnras, 475,
  2016

\bibitem[{{Dame} \& {Thaddeus}(1985)}]{1985ApJ...297..751D}
{Dame}, T.~M. \& {Thaddeus}, P. 1985, \apj, 297, 751

\bibitem[{{Dotson} {et~al.}(2010){Dotson}, {Vaillancourt}, {Kirby}, {Dowell},
  {Hildebrand}, \& {Davidson}}]{2010ApJS..186..406D}
{Dotson}, J.~L., {Vaillancourt}, J.~E., {Kirby}, L., {et~al.} 2010, \apjs, 186,
  406

\bibitem[{{Drozdovskaya} {et~al.}(2018){Drozdovskaya}, {van Dishoeck},
  {J{\o}rgensen}, {Calmonte}, {van der Wiel}, {Coutens}, {Calcutt},
  {M{\"u}ller}, {Bjerkeli}, {Persson}, {Wampfler}, \&
  {Altwegg}}]{drozdovskaya18}
{Drozdovskaya}, M.~N., {van Dishoeck}, E.~F., {J{\o}rgensen}, J.~K., {et~al.}
  2018, \mnras, 476, 4949

\bibitem[{{Dullemond} {et~al.}(2012){Dullemond}, {Juhasz}, {Pohl}, {Sereshti},
  {Shetty}, {Peters}, {Commercon}, \& {Flock}}]{2012ascl.soft02015D}
{Dullemond}, C.~P., {Juhasz}, A., {Pohl}, A., {et~al.} 2012, {RADMC-3D: A
  multi-purpose radiative transfer tool}, Astrophysics Source Code Library

\bibitem[{{Enoch} {et~al.}(2011){Enoch}, {Corder}, {Duch{\^e}ne}, {Bock},
  {Bolatto}, {Culverhouse}, {Kwon}, {Lamb}, {Leitch}, {Marrone}, {Muchovej},
  {P{\'e}rez}, {Scott}, {Teuben}, {Wright}, \&
  {Zauderer}}]{2011ApJS..195...21E}
{Enoch}, M.~L., {Corder}, S., {Duch{\^e}ne}, G., {et~al.} 2011, \apjs, 195, 21

\bibitem[{{Evans}(1999)}]{1999ARA&A..37..311E}
{Evans}, II, N.~J. 1999, \araa, 37, 311

\bibitem[{{Evans} {et~al.}(2015){Evans}, {Di Francesco}, {Lee}, {J{\o}rgensen},
  {Choi}, {Myers}, \& {Mardones}}]{2015ApJ...814...22E}
{Evans}, II, N.~J., {Di Francesco}, J., {Lee}, J.-E., {et~al.} 2015, \apj, 814,
  22

\bibitem[{{Fuller} {et~al.}(1995){Fuller}, {Lada}, {Masson}, \&
  {Myers}}]{1995ApJ...453..754F}
{Fuller}, G.~A., {Lada}, E.~A., {Masson}, C.~R., \& {Myers}, P.~C. 1995, \apj,
  453, 754

\bibitem[{{Garcia}(2011)}]{2011ppcd.book.....G}
{Garcia}, P.~J.~V. 2011, {Physical Processes in Circumstellar Disks around
  Young Stars}

\bibitem[{{Harsono} {et~al.}(2014){Harsono}, {J{\o}rgensen}, {van Dishoeck},
  {Hogerheijde}, {Bruderer}, {Persson}, \& {Mottram}}]{2014A&A...562A..77H}
{Harsono}, D., {J{\o}rgensen}, J.~K., {van Dishoeck}, E.~F., {et~al.} 2014,
  \aap, 562, A77

\bibitem[{{Hatchell} {et~al.}(1999){Hatchell}, {Fuller}, \&
  {Ladd}}]{1999A&A...344..687H}
{Hatchell}, J., {Fuller}, G.~A., \& {Ladd}, E.~F. 1999, \aap, 344, 687

\bibitem[{{Helou} \& {Walker}(1988)}]{1988iras....7.....H}
{Helou}, G. \& {Walker}, D.~W., eds. 1988, {Infrared astronomical satellite
  (IRAS) catalogs and atlases. Volume 7: The small scale structure catalog},
  Vol.~7, 1--265

\bibitem[{{Herbst} \& {van Dishoeck}(2009)}]{herbst09}
{Herbst}, E. \& {van Dishoeck}, E.~F. 2009, \araa, 47, 427

\bibitem[{{Hogerheijde} {et~al.}(1998){Hogerheijde}, {van Dishoeck}, {Blake},
  \& {van Langevelde}}]{1998ApJ...502..315H}
{Hogerheijde}, M.~R., {van Dishoeck}, E.~F., {Blake}, G.~A., \& {van
  Langevelde}, H.~J. 1998, \apj, 502, 315

\bibitem[{{Imai} {et~al.}(2016){Imai}, {Sakai}, {Oya}, {L{\'o}pez-Sepulcre},
  {Watanabe}, {Ceccarelli}, {Lefloch}, {Caux}, {Vastel}, {Kahane}, {Sakai},
  {Hirota}, {Aikawa}, \& {Yamamoto}}]{2016ApJ...830L..37I}
{Imai}, M., {Sakai}, N., {Oya}, Y., {et~al.} 2016, \apjl, 830, L37

\bibitem[{{Jacobsen} {et~al.}(2018){Jacobsen}, {J{\o}rgensen}, {van der Wiel},
  {Calcutt}, {Bourke}, {Brinch}, {Coutens}, {Drozdovskaya}, {Kristensen},
  {M{\"u}ller}, \& {Wampfler}}]{jacobsen18a}
{Jacobsen}, S.~K., {J{\o}rgensen}, J.~K., {van der Wiel}, M.~H.~D., {et~al.}
  2018, \aap, 612, A72

\bibitem[{{J{\o}rgensen}(2004)}]{2004A&A...424..589J}
{J{\o}rgensen}, J.~K. 2004, \aap, 424, 589

\bibitem[{{J{\o}rgensen} {et~al.}(2007){J{\o}rgensen}, {Bourke}, {Myers}, {Di
  Francesco}, {van Dishoeck}, {Lee}, {Ohashi}, {Sch{\"o}ier}, {Takakuwa},
  {Wilner}, \& {Zhang}}]{2007ApJ...659..479J}
{J{\o}rgensen}, J.~K., {Bourke}, T.~L., {Myers}, P.~C., {et~al.} 2007, \apj,
  659, 479

\bibitem[{{J{\o}rgensen} {et~al.}(2005){J{\o}rgensen}, {Bourke}, {Myers},
  {Sch{\"o}ier}, {van Dishoeck}, \& {Wilner}}]{2005ApJ...632..973J}
{J{\o}rgensen}, J.~K., {Bourke}, T.~L., {Myers}, P.~C., {et~al.} 2005, \apj,
  632, 973

\bibitem[{{J{\o}rgensen} {et~al.}(2012){J{\o}rgensen}, {Favre}, {Bisschop},
  {Bourke}, {van Dishoeck}, \& {Schmalzl}}]{2012ApJ...757L...4J}
{J{\o}rgensen}, J.~K., {Favre}, C., {Bisschop}, S.~E., {et~al.} 2012, \apjl,
  757, L4

\bibitem[{{J{\o}rgensen} {et~al.}(2002){J{\o}rgensen}, {Sch{\"o}ier}, \& {van
  Dishoeck}}]{2002A&A...389..908J}
{J{\o}rgensen}, J.~K., {Sch{\"o}ier}, F.~L., \& {van Dishoeck}, E.~F. 2002,
  \aap, 389, 908

\bibitem[{{J{\o}rgensen} {et~al.}(2004){J{\o}rgensen}, {Sch{\"o}ier}, \& {van
  Dishoeck}}]{2004A&A...416..603J}
{J{\o}rgensen}, J.~K., {Sch{\"o}ier}, F.~L., \& {van Dishoeck}, E.~F. 2004,
  \aap, 416, 603

\bibitem[{{J{\o}rgensen} {et~al.}(2016){J{\o}rgensen}, {van der Wiel},
  {Coutens}, {Lykke}, {M{\"u}ller}, {van Dishoeck}, {Calcutt}, {Bjerkeli},
  {Bourke}, {Drozdovskaya}, {Favre}, {Fayolle}, {Garrod}, {Jacobsen},
  {{\"O}berg}, {Persson}, \& {Wampfler}}]{2016A&A...595A.117J}
{J{\o}rgensen}, J.~K., {van der Wiel}, M.~H.~D., {Coutens}, A., {et~al.} 2016,
  \aap, 595, A117

\bibitem[{{J{\o}rgensen} {et~al.}(2009){J{\o}rgensen}, {van Dishoeck},
  {Visser}, {Bourke}, {Wilner}, {Lommen}, {Hogerheijde}, \&
  {Myers}}]{2009A&A...507..861J}
{J{\o}rgensen}, J.~K., {van Dishoeck}, E.~F., {Visser}, R., {et~al.} 2009,
  \aap, 507, 861

\bibitem[{{Ladd} {et~al.}(1991){Ladd}, {Adams}, {Casey}, {Davidson}, {Fuller},
  {Harper}, {Myers}, \& {Padman}}]{1991ApJ...366..203L}
{Ladd}, E.~F., {Adams}, F.~C., {Casey}, S., {et~al.} 1991, \apj, 366, 203

\bibitem[{{Lee} {et~al.}(2014){Lee}, {Hirano}, {Zhang}, {Shang}, {Ho}, \&
  {Krasnopolsky}}]{2014ApJ...786..114L}
{Lee}, C.-F., {Hirano}, N., {Zhang}, Q., {et~al.} 2014, \apj, 786, 114

\bibitem[{{Lindberg} {et~al.}(2014){Lindberg}, {J{\o}rgensen}, {Brinch},
  {Haugb{\o}lle}, {Bergin}, {Harsono}, {Persson}, {Visser}, \&
  {Yamamoto}}]{2014A&A...566A..74L}
{Lindberg}, J.~E., {J{\o}rgensen}, J.~K., {Brinch}, C., {et~al.} 2014, \aap,
  566, A74

\bibitem[{{Looney} {et~al.}(2000){Looney}, {Mundy}, \&
  {Welch}}]{2000ApJ...529..477L}
{Looney}, L.~W., {Mundy}, L.~G., \& {Welch}, W.~J. 2000, \apj, 529, 477

\bibitem[{{Milam} {et~al.}(2005){Milam}, {Savage}, {Brewster}, {Ziurys}, \&
  {Wyckoff}}]{2005ApJ...634.1126M}
{Milam}, S.~N., {Savage}, C., {Brewster}, M.~A., {Ziurys}, L.~M., \& {Wyckoff},
  S. 2005, \apj, 634, 1126

\bibitem[{{M{\"u}ller} {et~al.}(2016){M{\"u}ller}, {Belloche}, {Xu}, {Lees},
  {Garrod}, {Walters}, {van Wijngaarden}, {Lewen}, {Schlemmer}, \&
  {Menten}}]{2016A&A...587A..92M}
{M{\"u}ller}, H.~S.~P., {Belloche}, A., {Xu}, L.-H., {et~al.} 2016, \aap, 587,
  A92

\bibitem[{{Murillo} {et~al.}(2013){Murillo}, {Lai}, {Bruderer}, {Harsono}, \&
  {van Dishoeck}}]{2013A&A...560A.103M}
{Murillo}, N.~M., {Lai}, S.-P., {Bruderer}, S., {Harsono}, D., \& {van
  Dishoeck}, E.~F. 2013, \aap, 560, A103

\bibitem[{{Ohashi} {et~al.}(2014){Ohashi}, {Saigo}, {Aso}, {Aikawa},
  {Koyamatsu}, {Machida}, {Saito}, {Takahashi}, {Takakuwa}, {Tomida},
  {Tomisaka}, \& {Yen}}]{2014ApJ...796..131O}
{Ohashi}, N., {Saigo}, K., {Aso}, Y., {et~al.} 2014, \apj, 796, 131

\bibitem[{{Ossenkopf} \& {Henning}(1994)}]{1994A&A...291..943O}
{Ossenkopf}, V. \& {Henning}, T. 1994, \aap, 291, 943

\bibitem[{{Oya} {et~al.}(2016){Oya}, {Sakai}, {L{\'o}pez-Sepulcre}, {Watanabe},
  {Ceccarelli}, {Lefloch}, {Favre}, \& {Yamamoto}}]{2016ApJ...824...88O}
{Oya}, Y., {Sakai}, N., {L{\'o}pez-Sepulcre}, A., {et~al.} 2016, \apj, 824, 88

\bibitem[{{Oya} {et~al.}(2017){Oya}, {Sakai}, {Watanabe}, {Higuchi}, {Hirota},
  {L{\'o}pez-Sepulcre}, {Sakai}, {Aikawa}, {Ceccarelli}, {Lefloch}, {Caux},
  {Vastel}, {Kahane}, \& {Yamamoto}}]{2017ApJ...837..174O}
{Oya}, Y., {Sakai}, N., {Watanabe}, Y., {et~al.} 2017, \apj, 837, 174

\bibitem[{{Oya} {et~al.}(2018){Oya}, {Sakai}, {Watanabe}, {L{\'o}pez-Sepulcre},
  {Ceccarelli}, {Lefloch}, \& {Yamamoto}}]{2018ApJ...863...72O}
{Oya}, Y., {Sakai}, N., {Watanabe}, Y., {et~al.} 2018, \apj, 863, 72

\bibitem[{{Parker}(1988)}]{1988MNRAS.235..139P}
{Parker}, N.~D. 1988, \mnras, 235, 139

\bibitem[{{Parker} {et~al.}(1991){Parker}, {Padman}, \&
  {Scott}}]{1991MNRAS.252..442P}
{Parker}, N.~D., {Padman}, R., \& {Scott}, P.~F. 1991, \mnras, 252, 442

\bibitem[{{Sakai} {et~al.}(2014){Sakai}, {Sakai}, {Hirota}, {Watanabe},
  {Ceccarelli}, {Kahane}, {Bottinelli}, {Caux}, {Demyk}, {Vastel}, {Coutens},
  {Taquet}, {Ohashi}, {Takakuwa}, {Yen}, {Aikawa}, \&
  {Yamamoto}}]{2014Natur.507...78S}
{Sakai}, N., {Sakai}, T., {Hirota}, T., {et~al.} 2014, \nat, 507, 78

\bibitem[{{Sandford} \& {Allamandola}(1993)}]{1993ApJ...417..815S}
{Sandford}, S.~A. \& {Allamandola}, L.~J. 1993, \apj, 417, 815

\bibitem[{{Shirley} {et~al.}(2000){Shirley}, {Evans}, {Rawlings}, \&
  {Gregersen}}]{2000ApJS..131..249S}
{Shirley}, Y.~L., {Evans}, II, N.~J., {Rawlings}, J.~M.~C., \& {Gregersen},
  E.~M. 2000, \apjs, 131, 249

\bibitem[{{Skrutskie} {et~al.}(2006){Skrutskie}, {Cutri}, {Stiening},
  {Weinberg}, {Schneider}, {Carpenter}, {Beichman}, {Capps}, {Chester},
  {Elias}, {Huchra}, {Liebert}, {Lonsdale}, {Monet}, {Price}, {Seitzer},
  {Jarrett}, {Kirkpatrick}, {Gizis}, {Howard}, {Evans}, {Fowler}, {Fullmer},
  {Hurt}, {Light}, {Kopan}, {Marsh}, {McCallon}, {Tam}, {Van Dyk}, \&
  {Wheelock}}]{2006AJ....131.1163S}
{Skrutskie}, M.~F., {Cutri}, R.~M., {Stiening}, R., {et~al.} 2006, \aj, 131,
  1163

\bibitem[{{Strauss} {et~al.}(1990){Strauss}, {Davis}, {Yahil}, \&
  {Huchra}}]{1990ApJ...361...49S}
{Strauss}, M.~A., {Davis}, M., {Yahil}, A., \& {Huchra}, J.~P. 1990, \apj, 361,
  49

\bibitem[{{Tafalla} {et~al.}(2000){Tafalla}, {Myers}, {Mardones}, \&
  {Bachiller}}]{2000A&A...359..967T}
{Tafalla}, M., {Myers}, P.~C., {Mardones}, D., \& {Bachiller}, R. 2000, \aap,
  359, 967

\bibitem[{{Taquet} {et~al.}(2015){Taquet}, {L{\'o}pez-Sepulcre}, {Ceccarelli},
  {Neri}, {Kahane}, \& {Charnley}}]{2015ApJ...804...81T}
{Taquet}, V., {L{\'o}pez-Sepulcre}, A., {Ceccarelli}, C., {et~al.} 2015, \apj,
  804, 81

\bibitem[{{Terebey} {et~al.}(1984){Terebey}, {Shu}, \&
  {Cassen}}]{1984ApJ...286..529T}
{Terebey}, S., {Shu}, F.~H., \& {Cassen}, P. 1984, \apj, 286, 529

\bibitem[{{Tobin} {et~al.}(2012){Tobin}, {Hartmann}, {Chiang}, {Wilner},
  {Looney}, {Loinard}, {Calvet}, \& {D'Alessio}}]{2012Natur.492...83T}
{Tobin}, J.~J., {Hartmann}, L., {Chiang}, H.-F., {et~al.} 2012, \nat, 492, 83

\bibitem[{{T{\'o}th} {et~al.}(2014){T{\'o}th}, {Marton}, {Zahorecz},
  {Bal{\'a}zs}, {Ueno}, {Tamura}, {Kawamura}, {Kiss}, \&
  {Kitamura}}]{2014PASJ...66...17T}
{T{\'o}th}, L.~V., {Marton}, G., {Zahorecz}, S., {et~al.} 2014, \pasj, 66, 17

\bibitem[{{Xiang} \& {Turner}(1995)}]{1995ApJS...99..121X}
{Xiang}, D. \& {Turner}, B.~E. 1995, \apjs, 99, 121

\bibitem[{{Yamamura} {et~al.}(2010){Yamamura}, {Makiuti}, {Ikeda}, {Fukuda},
  {Oyabu}, {Koga}, \& {White}}]{2010yCat.2298....0Y}
{Yamamura}, I., {Makiuti}, S., {Ikeda}, N., {et~al.} 2010, VizieR Online Data
  Catalog, 2298

\bibitem[{{Yen} {et~al.}(2015){Yen}, {Takakuwa}, {Koch}, {Aso}, {Koyamatsu},
  {Krasnopolsky}, \& {Ohashi}}]{2015ApJ...812..129Y}
{Yen}, H.-W., {Takakuwa}, S., {Koch}, P.~M., {et~al.} 2015, \apj, 812, 129

\bibitem[Gaia Collaboration et al.(2018)]{gaia18} Gaia Collaboration, Brown, A.~G.~A., Vallenari, A., et al.\ 2018, \aap, 616, A1 

\bibitem[Hogerheijde et al.(1999)]{hogerheijde99} Hogerheijde, M.~R., van Dishoeck, E.~F., Salverda, J.~M., \& Blake, G.~A.\ 1999, \apj, 513, 350 

\bibitem[Kirk et al. (2013)]{kirk13} Kirk, H., Myers, P.~C., Bourke, T.~L., et al.\ 2013, \apj, 766, 115

\bibitem[Lindegren et al.(2018)]{lindegren18} Lindegren, L., Hern{\'a}ndez, J., Bombrun, A., et al.\ 2018, \aap, 616, A2 

\bibitem[Ortiz-Leon et al. (2018) ]{ortizleon18} Ortiz-Le\'{o}n, G.~N., Loinard, L., Dzib, S.~A., Kounkel, M., Galli, P.~A.~B. Tobin, J.~J., Evans, N.~J. II., Hartmann, L., Rodr{\'{i}}guez, L.~F., Bricen \~{o}, C., Torres, R.~M. \& Mioduszewski, A.~J., 2018, \apjl, submitted

\end{thebibliography}
\appendix
\section{Distance of L483 from Gaia-DR2 measurements}
\label{distance}

As noted in the introduction, L483 has traditionally been associated with the Aquila Rift/Serpens region toward which it appears in projection. Recent estimates of the distance toward the larger-scale Serpens and Aquila environment based on VLBA and Gaia Data Release 2 (DR2) parallax measurements \citep{ortizleon18}, place those clouds at a distance of 436$\pm$9~pc, at odds with the traditionally quoted distance toward L483 of 200~pc \citep{1985ApJ...297..751D}.

\begin{figure}
\resizebox{\hsize}{!}{\includegraphics{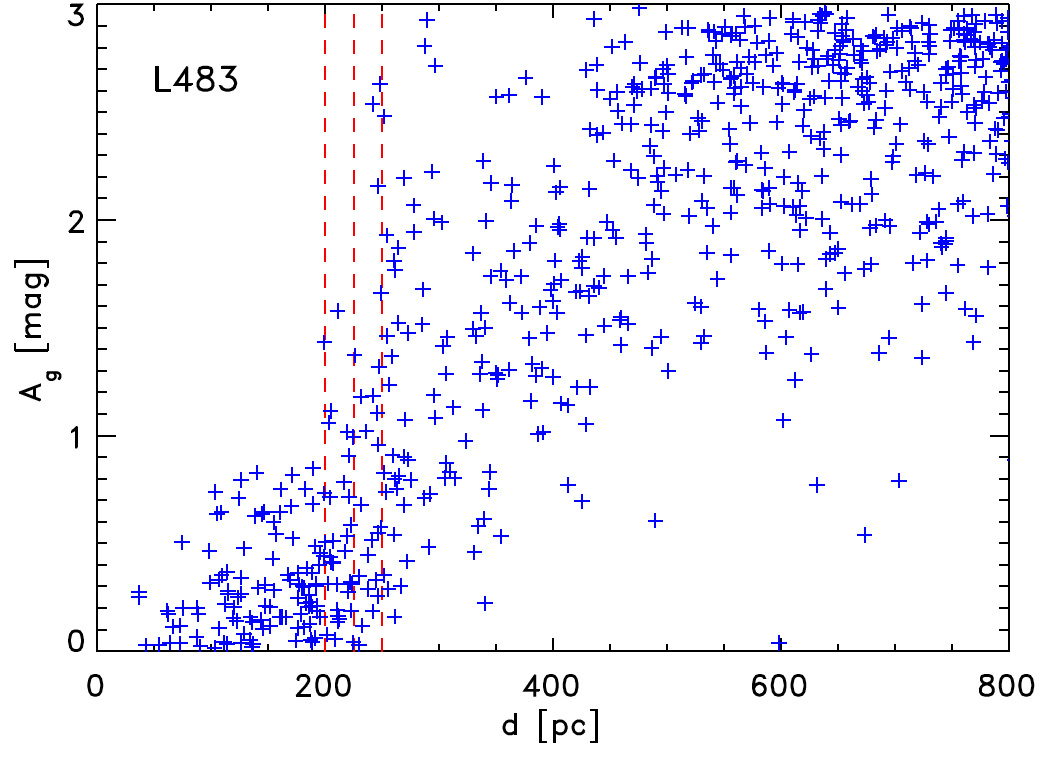}}
\caption{Extinction vs. distance for stars with reliable parallaxes from the Gaia-DR2 catalog within 1$^\circ$ of L483. The vertical dashed lines indicate distances of 200, 225, and 250~pc, where the extinction jumps significantly.}
\label{gaia_distance}
\end{figure}

To test whether the L483 is indeed also located at this farther distance, we extracted the extinctions and parallaxes for all stars within 1 degree of L483 from the Gaia-DR2 catalog \citep{gaia18}. Following \cite{ortizleon18}, we only selected stars with reliable parallax following the criteria described in \cite{lindegren18}. Fig.~\ref{gaia_distance} shows the extinction of each of these stars as function of distance. A sharp increase in the extinction of the stars at a distance of 200--250~pc is clearly seen, suggesting the presence of a cloud providing a significant amount of extinction at this distance. 
Millimeter-wavelength molecular line observations toward L483 (e.g., this paper) only show the presence of one cloud component at a LSR velocity of 5.0--5.5~km~s$^{-1}$. This suggests that L483 itself is at this nearer distance of 200--250~pc and thus not associated with the larger scale Serpens/Aquila cloud material.
Also, the LSR velocities of protostars in those larger-scale regions are found to be 7.5--8.5~km~s$^{-1}$ \citep[e.g.,][]{hogerheijde99,kirk13}, another indication that L483 and these clouds are not physically associated.

\section{Figures and tables}

\begin{table*}[h]
\centering
\caption{SED of L483. $^{a}$ 2MASS survey \citep{2006AJ....131.1163S}, $^{b}$ \citet{2014PASJ...66...17T}, $^{c}$\citet{2010yCat.2298....0Y}.}
\begin{tabular}{l*{8}{c}r}\toprule
\midrule
Wavelength [$\upmu$m] & Flux [Jy] & Flux uncertainty [Jy] & Reference \\
\hline
1.25 & 5.39$\times10^{-4}$ & 1.08$\times10^{-4}$ & 2MASS$^{a}$ \\
1.63 & 9.14$\times10^{-4}$ & 1.83$\times10^{-4}$ & 2MASS$^{a}$ \\ 
2.19 & 3.15$\times10^{-2}$ & 6.3$\times10^{-3}$ & 2MASS$^{a}$ \\
3.35 & 9.34$\times10^{-3}$ & 3.00$\times10^{-4}$ & WISE$^{b}$ \\
4.60 & 9.83$\times10^{-2}$ & 2.10$\times10^{-3}$ & WISE$^{b}$ \\ 
11.56 & 3.76$\times10^{-2}$ & 7.00$\times10^{-4}$ & WISE$^{b}$ \\
12.0 & 0.25 & 5$\times10^{-2}$ & \citet{1990ApJ...361...49S} \\ 
18.39 & 1.05 & 0.06 & Akari$^{c}$ \\
22.1 & 2.49 & 0.04 & WISE$^{b}$ \\
23.88 & 6.91 & 0.48 & WISE$^{b}$ \\
25.0 & 6.92 & 1.384 & \citet{1990ApJ...361...49S} \\
60.0 & 89.1 & 17.8 & \citet{1988iras....7.....H} \\
65 & 92.4 & 5.1 & AKARI$^{c}$ \\
90 & 87.4 & 4.0 & AKARI$^{c}$ \\
102 & 166.0 & 20.0 & \citet{1988iras....7.....H} \\
140 & 150.0 & 7.0 & AKARI$^{c}$ \\
160 & 156.0 & 7.0 & AKARI$^{c}$ \\
350 & 31.0 & 6.2 & \citet{2010ApJS..186..406D} \\
450 & 15.0 & 2.0 & \citet{2000ApJS..131..249S} \\
800 & 1.98 & 0.02 & \citet{2000ApJS..131..249S} \\
1100.0 & 0.64 & 0.02 & \citet{1995ApJ...453..754F} \\ 
\midrule
\bottomrule
\end{tabular}
\label{tab:sed}
\end{table*}
\begin{table*}
\begin{footnotesize}
  \caption{List of transitions of identified species predicted to be significant in modeling in Fig.~\ref{fig:spect_windows}.}
\label{transitions}
\begin{tabular}{lllll} \hline\hline
  Species & Transition  & Freq [GHz] & $\log_{10} A_{ul}$ [s$^{-1}$] & $E_u$ [K] \\
\hline
\multicolumn{5}{c}{\emph{Main targeted lines}}\\
HCN             & 4--3                              & 354.5055         & $-$2.68 & 43 \\
HCO$^+$         & 4--3                              & 356.7342         & $-$2.45 & 43 \\
CS              & 7--6                              & 342.8829         & $-$3.08 & 66 \\
H$^{13}$CN      & 4--3                              & 345.3398         & $-$2.69 & 41 \\
\hline                                                                                   
\multicolumn{5}{c}{\emph{Other assigned lines}}\\                                        
CH$_3$OH        & $23_{-4,20}-22_{-5,17}$           & 356.6272$^{*}$         & $-$4.09 & 728 \\ 
                & $18_{-8,10}-19_{-7,12}$           & 356.8749$^{*}$         & $-$4.31 & 718 \\ 
$^{13}$CH$_3$OH & $4_{0,4}-3_{-1,3}$                & 345.1326$^{*}$         & $-$4.08 & 36  \\ 
                & $4_{1,3}-3_{ 0,3}$                & 354.4459$^{*}$         & $-$3.89 & 44  \\ 
                & $3_{2,12}-12_{3,9}$               & 356.8738$^{*}$         & $-$4.10 & 244 \\
CH$_2$DOH       & $17_{2,16}-17_{1,17}$             & 342.9357               & $-$3.78 & 343 \\
                & $16_{2,14}-15_{3,13}$             & 345.3989               & $-$4.23 & 310 \\ 
                & $8_{7,1/2}-7_{7,0/1}$             & 356.7367$^a$           & $-$4.39 & 284 \\
                & $8_{6,2/3}-7_{6,1/2}$             & 356.7961$^{*}$         & $-$4.07 & 234 \\ 
                & $8_{2,7}-7_{2,6}$                 & 356.8997$^{\dagger}$   & $-$3.75 & 104 \\ 
                & $8_{5,3/4}-7_{5,2/3}$             & 356.9051$^{\dagger}$   & $-$3.94 & 193 \\ 
                & $8_{2,7}-7_{2,6}$                 & 356.9147               & $-$3.73 & 113 \\ 
                & $8_{4,4/5}-7_{4,3/4}$             & 356.9324$^{*}$         & $-$3.85 & 149 \\ 
CH$_3$OCH$_3$   & $8_{4,5} - 7_{3,4}$ (AE/EE) )     & 356.5753$^{\dagger}$   & $-$3.68 &  55 \\ 
                & $8_{4,5} - 7_{3,4}$ (AA)          & 356.5829$^{\dagger}$   & $-$3.68 &  55 \\ 
                & $8_{4,4} - 7_{3,4}$ (EE/EA) )     & 356.5868$^{\dagger}$   & $-$4.20 &  55 \\ 
                & $8_{4,5} - 7_{3,5}$ (EE)          & 356.7130$^{*}$         & $-$4.20 &  55 \\ 
                & $8_{4,4} - 7_{3,5}$ (EE)          & 356.7237$^{\dagger}$   & $-$3.83 &  55 \\
                & $8_{4,4} - 7_{3,5}$ (AA)          & 356.7245$^{\dagger}$   & $-$3.68 &  55 \\
CH$_3$OCHO      & $28_{ 6,23}-27_{ 6,22}$ (A) )     & 345.1480$^{*}$         & $-$2.31 & 452 \\ 
                & $28_{10,19}-27_{10,18}$ (E) )     & 345.2482               & $-$2.35 & 493 \\ 
                & $28_{13,15}-27_{13,14}$ (E) )     & 345.4610$^{\dagger}$   & $-$2.40 & 352 \\
                & $28_{13,15}-27_{13,14}$ (A) )     & 345.4670$^{*,\dagger}$ & $-$3.31 & 352 \\
                & $28_{ 9,19}-27_{ 9,18}$ (E) )     & 345.4732$^{\dagger}$   & $-$2.34 & 481 \\
                & $28_{13,16}-27_{13,15}$ (E) )     & 345.4866$^{*}$         & $-$2.40 & 352 \\
                & $28_{ 9,20}-27_{ 9,19}$ (A) )     & 345.5100$^{?}$         & $-$2.34 & 481 \\
                & $29_{13,16}-28_{13,15}$ (E) )     & 354.3487               & $-$2.45 & 556 \\
                & $33_{0/1,33}-32_{0/1,32}$ (E/A)   & 354.6078$^{*}$         & $-$2.55 & 293 \\
                & $28_{ 7,21}-27_{ 7,20}$ (E)       & 354.6287               & $-$2.29 & 461 \\ 
                & $29_{18,12}-28_{18,11}$ (A)       & 356.5398$^{*}$         & $-$2.55 & 471 \\
                & $29_{18,11}-28_{18,10}$ (E)       & 356.5559               & $-$2.55 & 471 \\
                & $29_{18,12}-28_{18,11}$ (E)       & 356.5663$^b$           & $-$2.55 & 471 \\
                & $29_{ 6,24}-28_{ 6,23}$ (A)       & 356.6869$^{?}$         & $-$2.36 & 469 \\ 
                & $29_{17,13/12}-28_{17,12/11}$ (A) & 356.7118$^c$           & $-$3.34 & 448 \\
                & $29_{17,12}-28_{17,11}$ (E)       & 356.7239$^d$           & $-$2.52 & 448 \\
                & $29_{17,13}-28_{17,12}$ (E)       & 356.7384$^a$           & $-$2.52 & 448 \\
                & $29_{10,19}-28_{10,18}$ (E)       & 356.7770$^?$           & $-$2.41 & 510 \\ 
                & $29_{11,19}-28_{11,18}$ (E)       & 356.8582$^?$           & $-$2.41 & 524 \\ 
                & $29_{16,14/13}-28_{16,13/12}$ (A) & 356.9287               & $-$3.32 & 426 \\
                & $29_{16,13}-28_{16,12}$ (E)       & 356.9363               & $-$2.50 & 426 \\
                & $29_{16,14}-28_{16,13}$ (E)       & 356.9545               & $-$2.50 & 426 \\ 
C$_2$H$_5$OH    & $7_{7,0/1}-6_{ 6, 0/1}, vt=0-1$   & 345.1739$^e$           & $-$3.60 & 140 \\
                & $21_{ 1,21}-20_{ 1,20}, vt=0-0$   & 345.2293               & $-$3.43 & 242 \\
                & $21_{ 1,21}-20_{ 1,20}, vt=1-1$   & 345.2954               & $-$3.44 & 246 \\
                & $21_{ 0,21}-20_{ 0,20}, vt=0-0$   & 345.3334$^f$           & $-$3.43 & 242 \\
                & $21_{ 0,21}-20_{ 0,20}, vt=1-1$   & 345.4082               & $-$3.44 & 246 \\
                & $20_{3,17}-19_{2,17}, vt=0-1$     & 354.3632$^{?}$         & $-$3.31 & 245 \\
CH$_3$CHO       & $18_{2, 16} - 17_{2, 15}$         & 354.4577$^{*,g}$       & $-$2.80 & 375 \\
                & $19_{0, 19} - 18_{0, 18}$         & 354.5254               & $-$2.80 & 377 \\
NH$_2$CHO       & $17_{0,17} - 16_{0,16}$           & 345.1813$^{*}$         & $-$2.52 & 152 \\
                & $16_{1,15} - 15_{1,14}$           & 345.3254               & $-$2.52 & 145 \\ 
                & $17_{2,16} - 16_{2,15}$           & 356.7138$^c$           & $-$2.48 & 167 \\ 
H$_2$CS         & $10_{ 0,10}- 9_{ 0, 9}$           & 342.9464               & $-$3.22 &  91 \\ 
HC$_3$N         & $39-38$                           & 354.6975$^{*}$         & $-$2.45 & 341 \\
SO$_2$          & $13_{ 2,12}-12_{ 1,11}$           & 345.3385$^f$           & $-$3.62 &  93 \\
                & $26_{ 9,17}-27_{ 8,20}$           & 345.4490               & $-$4.12 & 521 \\ 
                & $10_{ 4, 6}-10_{ 3, 7}$           & 356.7552               & $-$3.48 &  90 \\
\hline
\end{tabular}
\end{footnotesize}
\begin{scriptsize}

Notes: $^{\dagger}$Blended with nearby transitions of same species (blending taken into account in synthetic spectrum). $^{*}$ Transition shown in Fig.~\ref{fig:sel_lines}. $^{?}$Faint emission seen at frequency of transition but blend with other unidentified transitions possible. $^a$Blended with HCO$^+$ 4--3 transition. $^b$Blended with set of stronger CH$_3$OCH$_3$ transitions at 356.575--356.587 GHz. $^c$Blended with stronger CH$_3$OCH$_3$ transition at 356.7130 GHz. $^d$Blended with set of stronger CH$_3$OCH$_3$ transitions at 356.723--356.724 GHz. $^e$Blended with stronger NH$_2$CHO transition at 345.1813 GHz. $^f$Blended with H$^{13}$CN~$J$~=~4--3 transition. $^g$Synthetic spectrum cannot account for full observed line flux. Blending with unassigned line possible (see text). $^h$Blended with HCN~$J$~=~4--3 transition.
\end{scriptsize}
\end{table*}
\end{document}